\documentclass[sigconf]{acmart}

\usepackage{soul}
\usepackage{url}
\usepackage{graphicx}
\usepackage{subfigure}
\usepackage{amsmath}
\usepackage{amsthm}
\usepackage{booktabs}
\usepackage{algorithm}
\usepackage{algorithmic}
\usepackage{amsthm,amssymb}
\usepackage{mathrsfs}
\usepackage{multirow}
\usepackage{makecell}

\usepackage{bm}

\AtBeginDocument{%
  \providecommand\BibTeX{{%
    \normalfont B\kern-0.5em{\scshape i\kern-0.25em b}\kern-0.8em\TeX}}}

\copyrightyear{2020}
\acmYear{2020}
\setcopyright{acmlicensed}\acmConference[CIKM '20]{Proceedings of the 29th ACM International Conference on Information and Knowledge Management}{October 19--23, 2020}{Virtual Event, Ireland}
\acmBooktitle{Proceedings of the 29th ACM International Conference on Information and Knowledge Management (CIKM '20), October 19--23, 2020, Virtual Event, Ireland}
\acmPrice{15.00}
\acmDOI{10.1145/3340531.3411890}
\acmISBN{978-1-4503-6859-9/20/10}

\begin{document}
\fancyhead{}

\title{QSAN: A Quantum-probability based Signed Attention Network for Explainable False Information Detection}
\author{Tian Tian, Yudong Liu, Xiaoyu Yang, Yuefei Lyu, Xi Zhang}\authornote{Corresponding author} \author{Binxing Fang}
\email{{tiantian\_96727, yudong.liu, littlehaes, lvyuefei, 	zhangx, fangbx}@bupt.edu.cn}
\affiliation{\small
Key Laboratory of Trustworthy Distributed Computing and Service (MoE), Beijing University of Posts and Telecommunications, China}

\begin{abstract}
False information detection on social media is challenging as it commonly requires tedious evidence-collecting but lacks available comparative information. Clues mined from user comments, as the wisdom of crowds, could be of considerable benefit to this task. However, it is non-trivial to capture the complex semantics from the contents and comments in consideration of their implicit correlations. Although deep neural networks have good expressive power, one major drawback is the lack of explainability. In this paper, we focus on how to learn from the post contents and related comments in social media to understand and detect the false information more effectively, with explainability.  We thus propose a Quantum-probability based Signed Attention Network (QSAN) that integrates the quantum-driven text encoding and a novel signed attention mechanism in a unified framework. QSAN is not only able to distinguish important comments from the others, but also can exploit the conflicting social viewpoints in the comments to facilitate the detection. Moreover, QSAN is advantageous with its explainability in terms of transparency due to quantum physics meanings and the attention weights. Extensive experiments on real-world datasets show that our approach outperforms state-of-the-art baselines and can provide different kinds of user comments to explain why a piece of information is detected as false.
\end{abstract}

\maketitle

\section{Introduction}
The openness and convenience of social media have revolutionized the way of information dissemination. Meanwhile, it also reduces the cost of creating and sharing false information, which can be used by conspirators or cheaters to achieve their goals.
For example, during the 2016 U.S. presidential election, fake news accounted for nearly 6\% of all news consumption~\cite{grinberg2019fake}. Furthermore, it has affected stock markets~\cite{bollen2011twitter}, slowed responses during disasters~\cite{DBLP:conf/www/0003LKJ13}, and terrorist attacks~\cite{fisher2016pizzagate}. With the ever-increasing amounts of false information, it is of prominent importance to develop automatic detectors to mitigate the serious negative effects.


\begin{figure}
\begin{center}
\includegraphics[width=0.47\textwidth]{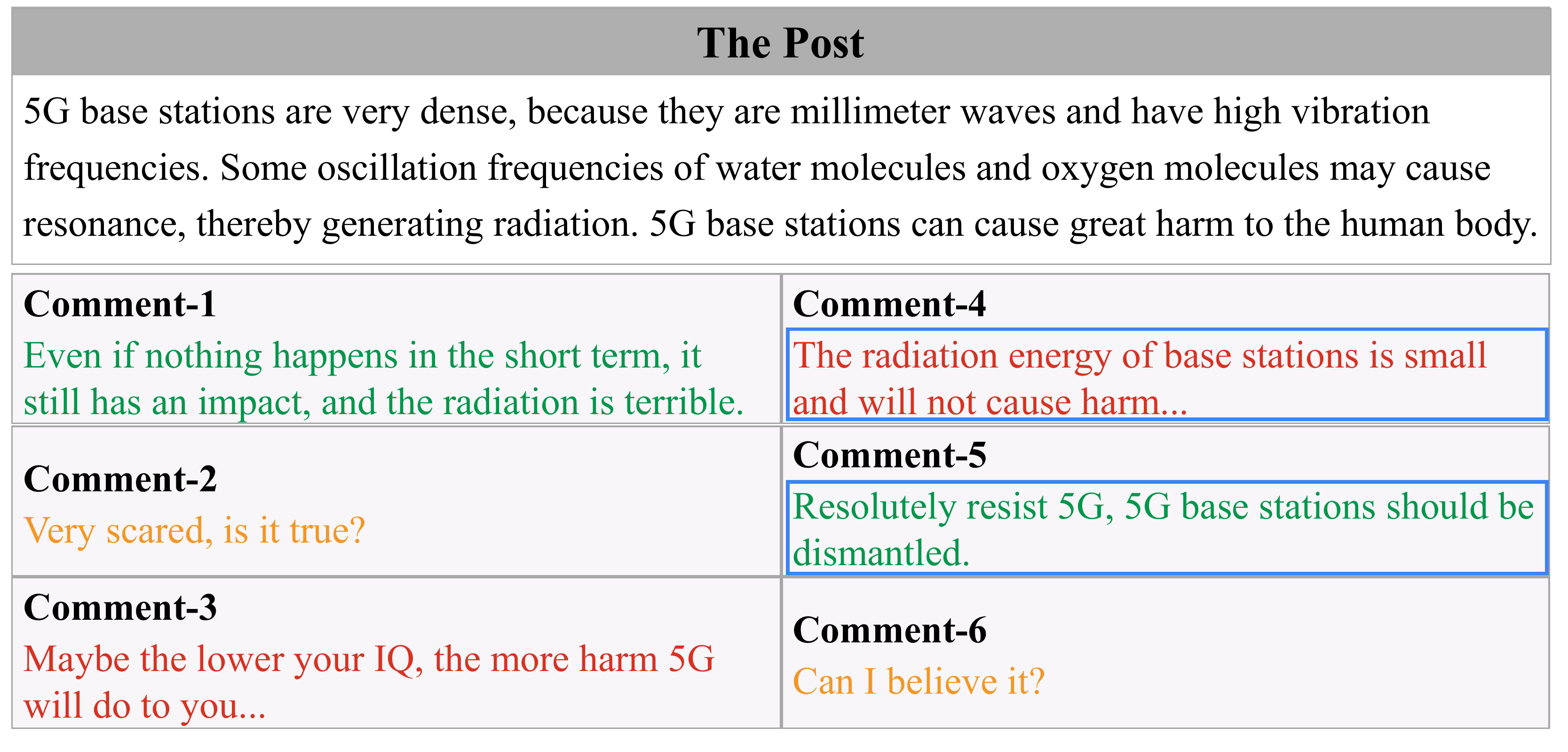}
\vspace{-0.3cm}
\caption{A post and its related comments on social media. The red and green comments express supporting and opposing viewpoints respectively. The comments in blue boxes provide more significant clues than the rest for false information detection.}
\label{fig1}
\vspace{-0.6cm}
\end{center}
\end{figure}


Thus far, various approaches, including both traditional learning~\cite{liu2015statistical,DBLP:conf/sdm/WuLHL17} 
and deep neural network-based (DNN) models~\cite{DBLP:conf/ijcai/MaGMKJWC16,tacchini2017some}, have been proposed to detect false information on social media. DNN models have achieved performance improvement over traditional ones due to their superior expressive power. However, the lack of explanations of "how the system works" and "why this decision is made", limit the faithfulness and practical utility of these models. Interpretable Machine Learning~\cite{DBLP:journals/cacm/DuLH20} would be an effective tool to mitigate these problems, but most of the explainable methods only provide one type of explainability. How to simultaneously ensure the transparency of the decision-making process in the model designing phase and provide meaningful explanations after a decision made is a challenging task. Although~\cite{DBLP:conf/kdd/ShuCW0L19} has made an initial attempt for explainable fake news detection with the co-attention mechanism, the diversity of the explanations is limited.

Existing models typically learn from text content for detecting false information~\cite{liu2015statistical, DBLP:conf/aaai/JinCZL16}, and thus the effectiveness of these models highly depends on the quality of text features or representations. Previous studies have shown that human language understanding~\cite{wang2016exploration,bruza2009there} exhibits certain non-classical phenomena (i.e. quantum-like phenomena), e.g., semantic contradictions and ambiguities~\cite{bruza2008entangling}. For example, the claim "two cars were reported stolen by the Groveton police yesterday" can be interpreted in different ways, that is, the police either reported or stole the two cars~\cite{al2011syntactic}. As the false information is commonly written with the intent to confuse or mislead readers, it is more likely to have such complex characteristics~\cite{wenzel2019verify}, which may not be fully captured by traditional text representation methods. Quantum-probability based framework can serve as a promising method to formulate the quantum-like phenomena and to better capture different levels and aspects of semantic units, which would be of considerable benefit for the task of false information detection. 


Moreover, detection merely with text content may suffer from the lack of available comparative information. To address this issue, detection models can take advantage of the user comments to learn richer discriminative features. However, this poses a technical challenge to explore various relationships between a post and its comments. Figure 1 illustrates a case where a piece of false information is more likely to be detected with the help of comments. Contributions of these comments can be explored from two perspectives, one is "importance" and the other is "stance". 
For example, comments in green and red express supporting and opposing viewpoints respectively, which indicate the stance of the comment, while comments in blue boxes provide more informative clues than the others indicating they may be more important.
Therefore, it is of considerable benefit if these meaningful comments are captured from noisy voices to enable better detection performance. However, prior studies either only account for the importance of comments with attention mechanism~\cite{DBLP:conf/kdd/ShuCW0L19}, or merely extract stances of comments with topic model~\cite{DBLP:conf/aaai/JinCZL16} or Recurrent Neural Network~\cite{DBLP:conf/www/MaGW18}. How to jointly consider the importance and stance of comments in one framework to enjoy the best of both worlds is less explored. Moreover, it would be a promising benefit if the model can provide these informative comments as explanations to help end-users understand the detection process and result.

In this work, we investigate how to identify a piece of false information with its related comments. We aim to optimize the accuracy and explainability in a joint and unified framework. To this end, we propose a Quantum-based Signed Attention Network (QSAN), where the text semantics can be modeled with a quantum-inspired text representation method, and the post-comment relationships can be captured with a novel signed attention mechanism. Specifically, We first follow the quantum probability and adopt a complex-valued network to encode post contents and user comments. Compared to traditional vector-based text representation methods, complex-valued networks have a richer representational capacity~\cite{DBLP:conf/naacl/LiWM19} and could facilitate in learning complex semantics from the false information. In addition, the network is transparent for that it is designed in alignment with quantum physics. We then devise a novel complex-valued signed attention mechanism, which can capture two types of relationships between the comments and post contents. One is whether a specific comment is important to the post, and the other is the viewpoint of a comment towards the post. With the help of the signed attention, important and opposing user comments can be selected to explain why a post is detected as fake. 
Therefore, QSAN is superior in explainability as it not only posses transparency in the decision-making process, but also provide helpful explanations after a decision is making.

To summarize, we make the following contributions:
\begin{itemize}
\item An effective quantum-based false information detection framework which provides explainability not only in presenting transparency in the decision-making process, but also giving meaningful explanations to the final detection results.
\item A quantum-based signed attention mechanism is devised to jointly learn different kinds of relationships between comments and posts.
\item Extensive experiments on real-world datasets demonstrate our proposal can outperform state-of-the-art baselines in terms of detection performance and model explainability.
\end{itemize}


\section{Related Work}
We briefly review the related works in three categories. 

\subsection{False Information Detection}
False information detection on social media aims at using relevant cues of suspected false information (such as text content, visual content, publisher's personal information, et.al) for identification~\cite{kumar2018false}. The early models were mainly based on text content, where statistical information~\cite{liu2015statistical} or hidden information (such as emotion or attitude)~\cite{DBLP:conf/aaai/JinCZL16} of the text content are extracted for classification. More recent deep learning methods have achieved better performance over traditional learning models. RNN-based methods~\cite{DBLP:conf/ijcai/MaGMKJWC16,DBLP:conf/acl/WongGM18} and CNN-based methods~\cite{ijcai2017-545} are proposed to capture high-level text semantics. In addition to text features, multimodal features~\cite{jin2016novel, khattar2019mvae} and social features~\cite{DBLP:conf/ijcai/YangLTLLZ20}
are also exploited by recent studies. Besides, a few works study some new related tasks such as false event detection~\cite{DBLP:conf/www/Ma0W19}, false picture detection~\cite{qi2019exploiting}, and how to detect the false information~\cite{guo2018rumor} at an early stage. However, these methods mostly lack the property of explainability.~\cite{DBLP:conf/kdd/ShuCW0L19} proposes an explainable fake news detection method with co-attention mechanism to select important comments. There are also some studies aiming to exploit the conflicting viewpoints to facilitate the detection~\cite{DBLP:conf/aaai/JinCZL16,DBLP:conf/www/MaGW18}, but it lacks works that jointly consider the importance and the stance of the comments, which is the main contribution of our work.                     
\subsection{Complex-valued Text Representation}
Complex-valued text representation has shown promising performance in NLP tasks such as text classification~\cite{DBLP:conf/www/WangLM019}, information retrieval~\cite{DBLP:series/irs/Melucci15} and question answering~\cite{DBLP:conf/naacl/LiWM19}. Complex embeddings are also integrated into NN-architectures such as Transformer~\cite{yang2019complex}, convolutional feed-forward networks and convolutional LSTMs~\cite{DBLP:conf/iclr/TrabelsiBZSSSMR18}, but they don't follow the quantum operations such as mixture and measurement. 
The emerging field of quantum cognition points out that human cognition~\cite{aerts2014quantum}, especially language understanding~\cite{wang2016exploration}, exhibits certain quantum-like phenomena, such as the uncertainty embodied in semantic contradictions and ambiguities. ~\cite{sadrzadeh2018sentence} shows that quantum mechanics theory has advantages in semantic representation.~\cite{bruza2009there} proposes a mathematical framework for modeling semantic feature spaces using quantum mechanics, and verifies that implicit semantic analysis and HAL (Hyperspace Analog to Language) models are essentially standardized Hilbert spaces.~\cite{DBLP:conf/www/WangLM019} applies quantum theory to text representation learning tasks.~\cite{DBLP:conf/naacl/LiWM19} implements a quantum probability-based model for the QA problem. However, to the best of our knowledge, no previous work has applied the quantum-inspired complex network for false information detection. Moreover, prior studies are unable to explicitly capture different kinds of correlations between texts for explainability.


Probabilistic data-based methods can also model the data uncertainty but are commonly used for the uncertainty of whether a record (or an attribute of a record) is correct or exists in the database~\cite{dalvi2009probabilistic,antova200910}. On the contrary, quantum-probability based methods have a wider application range such as text semantic modeling. In addition, the quantum probability is guided by quantum theory and uses complex-valued representations, which is also different from the probabilistic data which adopts real-valued representations.



\subsection{Attention Mechanism}
The attention mechanism in deep learning is essentially similar to the human biological systems~\cite{chaudhari2019attentive}. The core goal is to select the information that is more critical to the current task. Since the attention mechanism is proposed, it has been widely used in various tasks such as machine translation~\cite{bahdanau2014neural}
, text representation~\cite{kiela2018dynamic}, recommendation system~\cite{ying2018sequential}, etc. 
~\cite{bahdanau2014neural} applies attention mechanisms for machine translation to select critical input vocabulary to the target translation result. In the task of natural language processing, the popularity of Transformer~\cite{vaswani2017attention} has also drawn increasing attention to self-attention.~\cite{DBLP:conf/nips/LuYBP16} proposes co-attention mechanism for visual question answering, trying to learn attention weights for images and questions at the same time. For false information detection, the attention mechanism can help us distinguish the confusing input, select more key features, and improve the explainability of the model. Although existing false information detection methods have used the attention mechanism~\cite{guo2018rumor,DBLP:conf/kdd/ShuCW0L19}
, they only consider the importance of the input. In contrast, our proposed signed attention mechanism can jointly consider the importance and stance correlations between comments and posts, which would be beneficial to applications that require capturing richer semantic relationships. 


\section{Preliminary}
Here we briefly introduce the theoretical framework called Semantic Hilbert Space defined on a complex-valued vector space. It is able to formulate quantum-like phenomena in language understanding and model different levels of semantic units in a unified space~\cite{DBLP:conf/www/WangLM019}.

In Semantic Hilbert Space, following the standard  \emph{Dirac Notation} of Quantum Theory, the unit vector and its transpose are denoted as a ket $\vert\mu\rangle$ and a bra $\langle\mu\vert$. The inner and outer product of two unit vectors $\vec{u}$ and $\vec{v}$ are denoted as $\langle u\vert v\rangle$ and $\vert u\rangle\langle v\vert$ respectively.

Sememes are the minimal non-separable semantic units of word meanings in language universals~\cite{goddard1994semantic}. In Semantic Hilbert Space, the set of sememes, $\{e_j\}^n_{j=1}$ are modeled as \emph{basis states}, and they are a set of orthogonal basis, $\{\vert e_j \rangle \}^n_{j=1}$, which is the basis for representing any quantum state.

A word is regarded as the \emph{superposition state}, which is represented by a unit-length vector. A word $w$ can thus be written as a linear combination of \emph{basis states} for sememes:
\begin{align}\small
\vert w\rangle=\sum\limits_{j=1}^n r_j e^{i\phi_j}\vert e_j\rangle
\end{align}
where the complex-valued weight $r_j e^{i\phi_j}$ denotes how much the meaning of word $w$ is associated with the sememe $\vert e_j\rangle$, $i$ is the imaginary unit. Here $\{r_j\}^n_{j=1}$ are non-negative real-valued amplitudes satisfying $\sum\limits_{j=1}^n r_j^2 = 1$ and $\phi_j\in[- \pi,\pi]$ are the phases. 

A word composition (bags of words, sentences, etc.) is considered as a \emph{mixed system} of words, based on the concept of the \emph{quantum mixture}, represented by a density matrix $\rho$:
\begin{align}\small
\rho=\sum\limits_{i} p_{i} \vert w_i\rangle\langle w_i\vert
\end{align}
where $\vert w_i\rangle$ denotes the state of $i$-th word and $p_i$ is the classical probability of the state $\vert w_i\rangle$ with $\sum\limits_{i} p_i=1$. It determines the contribution of the word $w_i$ to the overall semantics.

As a non-classical probability distribution, a density matrix carries rich information of a sequence of words. Its diagonal elements are real and form a classical distribution of sememes, while its off-diagonal elements are complex-valued and encode interactions between sememes, thereby forming interference between words. To capture the high-level features, a set of measurements can be built, associated with the measurement projectors $\{P_i\}^Z_{i=1}$. Given a density matrix $\rho$ and a measurement projector $P_i$, the measurement results can be obtained through:
\begin{align}\small
q_i=tr(P_i\rho)
\end{align} 

Here, measurement states are only considered as pure states, i.e. $P_i=\vert v_i\rangle\langle v_i\vert$, where $\vert v_i\rangle$ could be any pure state (not limited to a specific word $w$). For a specific task, $\{\vert v_i\rangle\}_Z$ would serve as trainable parameters so that the most suitable measurements can be determined automatically by the training data. And these measurement operators collapse the mixed system into $Z$ different states. 
For example, a density matrix $\rho$ is measured by a measurement factor $P_i$ to obtain a real value $q_i$, which represents the probability that the density matrix collapses to the $i$-th state. So given a set of measurement factors, a set of $q_i$ can be obtained, which are then connected to obtain a real-valued vector as the output of measurement operations, where each dimension of the vector represents the probability of being in a certain state.

Our framework requires multiplication between two matrices $\bm{W}$ and $\bm{A}$ in the complex field, whose result $\bm{B}$ can be obtained as:
\begin{equation}\small
    \begin{aligned}
        \bm{B}^{re}&=\bm{W}^{re}\bm{A}^{re}-\bm{W}^{im}\bm{A}^{im}\\    
        \bm{B}^{im}&=\bm{W}^{re}\bm{A}^{im}+\bm{W}^{im}\bm{A}^{re}
    \end{aligned}
\end{equation}
where $\bm{X}^{re}$, $\bm{X}^{im}$ ($\bm{X}$ = $\bm{W}$, $\bm{A}$, or $\bm{B}$) represent the real and imaginary parts of the matrix respectively, that is, $\bm{X}=\bm{X}^{re}+i*\bm{X}^{im}$.


\section{Problem Statement}
We study the problem of detecting false information by using the contents of the post and its related comments. We assume that a post $I$ contains $N$ sentences $S_i$, $I = \{S_i\}^N_{i=1}$, and its comment set is denoted as $R$, consisting of $T$ comments, $R = \{C_i\}^T_{i=1}$. We treat the false information detection problem as a binary classification task. By learning the semantic representations of the post and comments texts and capturing the semantic relationships between them, we aim to get the label of the post $y \in \{0, 1\}$, where $y=1$ indicates the post is false, otherwise $y=0$. Therefore, the problem can be abstracted as learning a mapping function $f$, $f(I, R)\rightarrow y$, from the content of the posts and comments to the labels to maximize prediction accuracy.

\section{QSAN: A Quantum-probability based Signed Attention Network}

\subsection{Overview}
Our framework QSAN is designed based on quantum probability and attention mechanism over Semantic Hilbert Space. QSAN mainly consists of three components (see Figure 1): (1) a quantum probability-driven sentence and comment encoder (complex embedding and mixture component); (2) a quantum-based signed attention component; (3) a semantic measurement component. 

\begin{figure*}
\begin{center}
\includegraphics[width=0.95\textwidth]{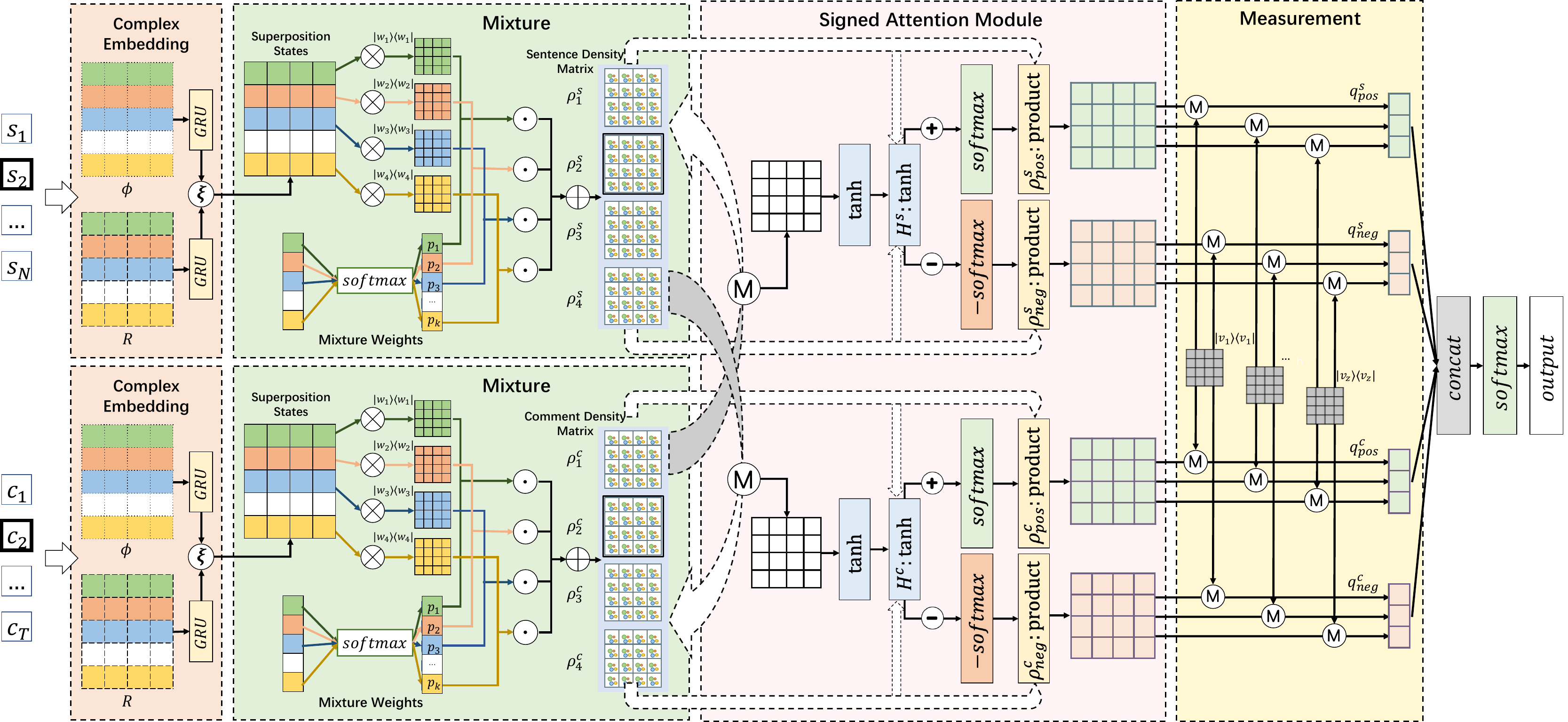}
\caption{QSAN consists of three components: (1) a quantum probability-driven sentence and comment encoder (complex embedding and mixture component); (2) a quantum-based signed attention component; (3) a semantic measurement component. } \label{fig1}
\end{center}
\vspace{-0.4cm}
\end{figure*}


The data flows in QSAN as follows. We first feed the post contents (denoted as $\{S_i\}^N_{i=1}$ indicating sentences) and user comments (denoted as $\{C_i\}^T_{i=1}$) to the quantum probability-driven encoder to obtain their complex-valued density matrices. They can be seen as the encoded representations of the sentences and comments in Semantic Hilbert Space, which are then served as inputs to the quantum-based signed attention component (Sec. 5.2). The quantum-based signed attention module is equipped with two-channel softmax functions, with the same input but different outputs. Then for each comment, its importance in supporting the post is learned as the attention weight in one channel (named as "+softmax"), and similarly, the importance in opposing the post is learned in the other channel (named as "-softmax"). In this way, the signed attention mechanism can capture the importance and stances of comments at the same time (Sec. 5.3), and the attention weights can help to select informative comments as explanations. Finally, after the signed attention component, we obtain the complex-valued feature density matrices, which are then fed into the semantic measurement component to obtain the high-level real-valued semantic abstractions, which can be seen as a dimension reduction step. Then a fully connected layer is applied to connect the high-level abstractions of the sentences and comments to obtain the final classification labels (Sec. 5.4).


\subsection{Quantum Probability-Driven Sentences and Comments Encoder}
Inspired by~\cite{DBLP:conf/www/WangLM019}, we develop the quantum probability-driven encoder over the Semantic Hilbert Space, aiming to capture the non-linear representation of each sentence and comment. Thus, the semantics of sentences and comments are expressed more comprehensively, and better prerequisites are provided for signed attention modules to extract the semantic relationship between them.

\subsubsection{Complex Embedding for Words}
In the Semantic Hilbert Space, each level of the semantic unit (sememe, word or word composition) is represented as complex-valued vector. For each word $w$ in a sentence (or in a comment, processed in the same way as in a sentence, thus omitted in the remainder), it is regarded as the \emph{superposition state} of sememes, $\{e_j\}^d_{j=1}$, represented as $\vert w\rangle=[ r_1 e^{i\phi_1},r_2 e^{i\phi_2},...,r_d e^{i\phi_d} ]^T$. The complex-valued representation allows a non-linear combination of amplitude $r_j$ and phase $\phi_j$, thereby indicating a non-linear combination of semantics. Suppose two words $w_1$ and $w_2$ are with weights $r_j^{(1)} e^{i\phi_j^{(1)}}$ and $r_j^{(2)} e^{i\phi_j^{(2)}}$ for the $j^{th}$ dimension (i.e., the $j^{th}$ sememe $e_j$), the composition of the two words for $e_j$ is computed as:
\begin{equation}\small
    \begin{aligned}
        r_j e^{i\phi_j}=r_j^{(1)} e^{i\phi_j^{(1)}}+r_j^{(2)} e^{i\phi_j^{(2)}}
    \end{aligned}
\end{equation}
where
\begin{equation}\small
    \begin{aligned}
        r_j&=\sqrt{\vert r_j^{(1)} \vert^2 + \vert r_j^{(2)} \vert^2 + 2 r_j^{(1)} r_j^{(2)} cos( \phi_j^{(1)} - \phi_j^{(2)} )}\\
        \phi_j&=arctan(\frac{r_j^{(1)} sin(\phi_j^{(1)})+r_j^{(2)} sin(\phi_j^{(2)})}{r_j^{(1)} cos(\phi_j^{(1)})+r_j^{(2)} cos(\phi_j^{(2)})})
    \end{aligned}
\end{equation}

Following the embedding initialization method in~\cite{DBLP:conf/www/WangLM019}, we use a pre-trained dictionary (Glove for English and the embedding in~\cite{P18-2023}\cite{qiu2018revisiting} for Chinese) to obtain the embedding of the word amplitude, and at the same time, randomly generate the phase embedding. They are fed into GRU to capture the contextual relevance of the word in the sentence to obtain the initial amplitude embedding $[r_1,r_2,...,r_d]^\top$ and the initial phase embedding $[\phi_1,\phi_2,...,\phi_d]^\top$ of the word complex-valued vector. Here $d$ represents the dimension of the word vector, which is fixed for all words, and can also be understood as the number of basis states or sememes to build the word. Finally, complex-valued representation of each word is expressed as: $\vert \textbf{w}\rangle=[ r_1 e^{i\phi_1},r_2 e^{i\phi_2},...,r_d e^{i\phi_d} ]^\top$.

\subsubsection{Mixture Component for Sentences and Comments}
Based on the concept of the quantum mixture, a sentence is considered as a mixed system of words, represented by a complex-valued quantum density matrix. The words in a sentence are mixed according to the mixture probability to obtain a mixture density matrix. With the mixture probability of word $\textbf{w}_i$, denoted as $p(\textbf{w}_i)$, which is a real number, the sentence mixture density matrix is obtained as:
\begin{align}\small
\boldsymbol{\rho}=\sum\limits_{i=1}^m p(\textbf{w}_i) \vert \textbf{w}_i\rangle\langle \textbf{w}_i\vert
\end{align}
where $m$ represents the number of words in the sentence, and the summation operation is similar to Eq. (5).

Note that we have $\sum\limits_{i=1}^{m} p(\textbf{w}_i)=1$ to guarantee the unit trace length for a density matrix. We can enforce this property with a softmax operation to normalize word-dependent weights. The probability $p_i$ is not static but initialized randomly and adaptively updated in our framework. With the mixture component, sentences $\{S_i\}^N_{i=1}$ and comments $\{C_i\}^T_{i=1}$ are expressed as two sets of mixture density matrices: $\{\bm{\rho}^s_i\}^N_{i=1}$, $\{\bm{\rho}^c_i\}^T_{i=1}$ with complex values respectively, where $\bm{\rho}^s_i\in\mathbb{C}^{d*d}$, $\bm{\rho}^c_i\in\mathbb{C}^{d*d}$. We thus obtain the complex representations of sentences and comments and would feed them into the signed attention module described in the next subsection.

\subsection{Quantum-based Signed Attention Component}



In this work, we assume that each comment may present two views towards the post, i.e., supporting and opposing, at the same time. The role of the signed attention module is to capture the importance of the comment in each view, aiming to facilitate the false information detection task and provide meaningful explanations simultaneously.
We argue that traditional attention mechanisms such as  co-attention~\cite{DBLP:conf/kdd/ShuCW0L19} may not be sufficient to model such complex relationships among comments and posts, e.g., the comment opposing a post. Specifically, according to~\cite{vaswani2017attention}, an attention function can be described as mapping a query and a set of key-value pairs to an output. The output is computed as a weighted sum of the values, where the weight is computed by a compatibility function of the query with the corresponding key. In practice, a common implementation way is to compute the dot products of the query with the keys, and apply a softmax function to obtain the weights. The limitation is that the softmax function would treat the negative correlation between query and key as insignificance, which may not hold in many applications. 

For example, with the compatibility function, the resulting compatibility vector among a query and four corresponding keys (before softmax) is $[0.8, 0.2, -0.1, -0.6]$. Feeding it into softmax, the normalized weighting vector would be [0.45, 0.25, 0.18, 0.11]. It can be observed that the value associated with ``-0.6" in the compatibility vector would make the least contribution to the output. However, as the negative correlation, ``-0.6'' may indicate the two vectors pointing opposing directions, which can be understood as the comment opposing the post. It should be taken into account in our task.


\subsubsection{Signed Attention with Real Values} 
From the perspective of the ``opposing" relationship, in the above example, the importance order should be reversed, i.e., ``-0.6'' and ``0.8'' would be the most and least important factors respectively. To achieve this goal, we take the opposite values, then feed $[-0.8, -0.2, 0.1, 0.6]$ into a softmax function, and output the opposite of resulting weights. The output is $[-0.11, -0.20, -0.26, -0.43]$. We name this operation as ``-softmax". To make a distinct difference, the traditional softmax operation is called ``+softmax". To fully capture relationships from different perspectives, we utilize the weights from ``+softmax'' and ``-softmax" respectively to obtain the weighted sum of the values. We then feed the concatenation of the two vectors into the fully-connected layer for classification. We implement this signed attention idea in real values to replace the co-attention component, and achieve significant performance improvements (see the experimental Sec. 6.4). The complex-valued version of the signed attention mechanism designed for our QSAN is described as follows.

\subsubsection{Signed Attention in Semantic Hilbert Space}
Following the above intuition, we first obtain the affinity matrix and attention maps (concepts borrowed from co-attention~\cite{DBLP:conf/nips/LuYBP16}) from representations of sentences and comments, and then apply ``+softmax'' and  ``-softmax'' to obtain the weights from different perspectives. All the operations are in complex values.

Since each sentence and comment is denoted as a mixed state of words, i.e., the density matrix, we obtain the affinity matrix in Semantic Hilbert Space by applying the operation like the quantum measurement in Eq. (3). Inspired by co-attention, this operation can be understood as calculating the similarity between the mixed states, which is similar to the measurement operation over Semantic Hilbert Space. Therefore, in this paper, the density matrices of sentences and comments are regarded as each other’s measurement projectors to obtain the affinity matrix $\bm{L}$:
\begin{align}\small
\bm{M}_{ij}&=tr(\bm{\rho}^s_i\bm{\rho}^c_j)\\
\bm{L}&=tanh(\bm{M})
\end{align}
where $\bm{\rho}^s\in\mathbb{C}^{N\times d\times d}$, $\bm{\rho}^c\in\mathbb{C}^{T\times d\times d}$. Since the result of the quantum measurement is a real value, $\bm{L}$, $\bm{M}\in\mathbb{R}^{N\times T}$. 

Similar to co-attention, the affinity matrix $\bm{L}$ is considered as a feature and used to produce sentence and comment attention maps:
\begin{equation}\small
    \begin{aligned}
        \bm{H}^{s}&=ctanh(\bm{\rho}^s\bm{W}^s+\bm{L}(\bm{\rho}^c\bm{W}^c))\\
        \bm{H}^{c}&=ctanh(\bm{\rho}^c\bm{W}^c+\bm{L}^\top(\bm{\rho}^s\bm{W}^s))
    \end{aligned}
\end{equation}
where $\bm{W}^{s}$, $\bm{W}^{c}\in\mathbb{C}^{d\times d \times k}$, $\bm{H}^{s}\in\mathbb{C}^{N\times k}$, $\bm{H}^{c}\in\mathbb{C}^{T\times k}$. The $ctanh$ is the $tanh$ activation function for complex values: $ctanh(\textbf{a}) = tanh(\textbf{a}^{re})+i*tanh(\textbf{a}^{im})$, so is the $csoftmax$.

We then feed the attention maps $\bm{H}^{s}$, $\bm{H}^{c}$ into both ``+softmax'' and  ``-softmax'' functions to obtain the corresponding weights in two views.
The calculation process is: 

\begin{equation}\small
    \begin{aligned}
        \bm{a}^{s_{pos}}&=csoftmax(\bm{W}^{s_{pos}}(\bm{H}^{s})^\top)\\
        \bm{a}^{c_{pos}}&=csoftmax(\bm{W}^{c_{pos}}(\bm{H}^{c})^\top)
    \end{aligned}
\end{equation}
\begin{equation}
    \begin{aligned}
        \bm{a}^{s_{neg}}&=-csoftmax(-\bm{W}^{s_{neg}}(\bm{H}^{s})^\top)\\
        \bm{a}^{c_{neg}}&=-csoftmax(-\bm{W}^{c_{neg}}(\bm{H}^{c})^\top)
    \end{aligned}
\end{equation}
where $\bm{W}^{s_{pos}}$, $\bm{W}^{s_{neg}}$, $\bm{W}^{c_{pos}}$, $\bm{W}^{c_{neg}}\in\mathbb{C}^{1*k}$, and we thus obtain $\bm{a}^{s_{pos}}$, $\bm{a}^{s_{neg}}\in\mathbb{C}^{1*N}$ and $\bm{a}^{c_{pos}}$, $\bm{a}^{c_{neg}}\in\mathbb{C}^{1*T}$.

Based on the weights, complex-valued feature matrices of sentences and comments can be computed as:
\begin{equation}\small
    \begin{aligned}
        \bm{\rho}^s_{pos}&=\sum\limits_{i=1}^N \bm{a}_i^{s_{pos}} \bm{\rho}^s_i\\
        \bm{\rho}^c_{pos}&=\sum\limits_{i=1}^T \bm{a}_i^{c_{pos}} \bm{\rho}^c_i
    \end{aligned}
    \qquad
    \begin{aligned}
        \bm{\rho}^s_{neg}&=\sum\limits_{i=1}^N \bm{a}_i^{s_{neg}} \bm{\rho}^s_i\\
        \bm{\rho}^c_{neg}&=\sum\limits_{i=1}^T \bm{a}_i^{c_{neg}} \bm{\rho}^c_i
    \end{aligned}
\end{equation}
where $\bm{\rho}^s_{pos}, \bm{\rho}^s_{neg}, \bm{\rho}^c_{pos}, \bm{\rho}^c_{neg}\in\mathbb{C}^{d*d}$.



With the help of the singed attention mechanism, the obtained complex feature matrices of sentences and comments have taken their correlations into account. Different from the traditional co-attention mechanism that only considers the positive correlations, the signed attention module is added with a new channel called "-softmax" to capture the negative correlations as well. Each channel would produce a distinct feature matrix in the complex field, which would be fused to obtain a high-level representation in the real field for classification, which would be described in the next subsection.

Besides the feature matrices, the attention weight $a^{c_{pos}}$ (resp. $a^{c_{neg}}$) in the signed attention module would indicate the importance of a comment in supporting (resp. opposing) the post, providing informative comments for explanations. We would motivate this idea with a real case and describe it in detail in Sec. 6.5.

\subsection{Semantic Measurement Component}
We then describe how to fuse the feature matrices of sentences and comments obtained from the signed-attention module to produce the final output. 
Each feature density matrix contains the semantic possibility of the post (or the comments) under the false information detection task. We apply the measurement operation proposed in Eq. (3) to automatically learn a set of measurement operators through the labels, to make the semantics in the feature density matrix collapse to some certain states with probability, and the most suitable state is selected by the fully connected layer to get the final output.
Specifically, a set of rank-one measurement projectors $\{\vert \textbf{v}_i\rangle\langle \textbf{v}_i\vert\}^Z_{i=1}$ are used to measure the semantic of the feature matrices. After each measurement, we can obtain the probability $q_i = tr (\bm{\rho}\vert \textbf{v}_i\rangle\langle \textbf{v}_i\vert)$ corresponding to each state, and finally we will get a measurement state vector, $\textbf{q} = [q_1, q_2, ..., q_Z]$, as a high-level representation of the feature matrix in the real field.

Then all the $\textbf{q}$ of $\bm{\rho}^s_{pos}, \bm{\rho}^s_{neg}, \bm{\rho}^c_{pos}$ and $\bm{\rho}^c_{neg}$ are concatenated to form a real-valued vector and get the final output $\widehat{y}$ through a fully connected layer, where $\widehat{y}$ is [1, 0] when it is a piece of false information and [0, 1] otherwise. We adopt the cross-entropy loss function which is widely used in binary classification problems.

\section{EXPERIMENTS}
In this section, we conduct experiments to evaluate the effectiveness of the proposed QSAN framework. 


\begin{table}
    \small
    \caption{The Statistics of Datasets}
    \label{tab:datasets}
    \setlength{\tabcolsep}{1.8mm}{
    \begin{tabular}{c|c|c|c|c}
        \toprule
        \textbf{Datasets} & 
        \textbf{\# Posts} & 
        \textbf{\# False Posts}& \textbf{\# True Posts} & 
        \textbf{\# Comments} \\
        \midrule
        \textbf{FNews}  &1901 &1199 &702 &81385\\
        \textbf{CED}  &3382 &1538 &1844
        &161977\\
        \bottomrule
    \end{tabular}}

\end{table}


\subsection{Datasets}
We utilize two real-world false information datasets with both posts and comments. One is a fake news dataset in English called FakeNewsNet (i.e., FNews for short)~\cite{shu2018fakenewsnet} collected from two fact-checking platforms: GossipCop and PolitiFact, which includes the related retweets crawled on Twitter. The other is a rumor dataset in Chinese called CED~\cite{song2018ced}, which is collected from Weibo. As our model attempts to explore different kinds of relationships between the comments and the post, for preprocessing, we filter out the posts with less than 3 comments. We also delete the short comments with less than 10 characters and the duplicate comments. The statistics of the datasets are shown in Table 1. 




\subsection{Baselines}

\begin{table*}
\linespread{1.05}
\centering
\vspace{-0.5cm}
\caption{\textbf{The performance of baselines and QSAN}}
\label{tab:perfomance}
\fontsize{9}{14}\selectfont
\scriptsize
\scalebox{1.1}{
\begin{tabular}{c| c| c| c| c| c| c| c| c| c| c| c| c| c}
\toprule
{\textbf{Datasets}} & {\textbf{Metric}}&  {\textbf{LIWC}}& {\textbf{TextCNN}}& {\textbf{LSTM}}& {\textbf{BERT}}&{\textbf{HAN}}& {\textbf{Transformer}}& {\textbf{dEFEND}}&{\textbf{EANN}}& {\textbf{RumorGAN}}&  {\textbf{RVNN$_{BU}$}}& {\textbf{RVNN$_{TD}$}}&{\textbf{QSAN}}\cr
\midrule
\multirow{4}*{\textbf{CED}}&{Acc.}&{0.643}&{0.802}&{0.781}&{0.913}&{0.865}&{0.896}&{0.915}&{0.866}&{0.868}&{0.914}&{0.848}&{\textbf{0.957}}\\

&{Prec.}&{0.688}&{0.818}&{0.817}&{0.913}&{0.850}&{0.893}&{0.920}&{0.867}&{0.867}&{0.912}&{0.844}&{\textbf{0.968}}\\

&{Rec.}&{0.680}&{0.721}&{0.687}&{0.905}&{0.816}&{0.893}&{0.891}&{0.866}&{0.867}&{0.912}&{0.847}&{\textbf{0.938}}\\

&{$F_1$}&{0.683}&{0.766}&{0.746}&{0.909}&{0.832}&{0.893}&{0.905}&{0.866}&{0.867}&{0.912}&{0.845}&{\textbf{0.953}}\\

\midrule

\multirow{4}*{\textbf{FNews}}&{Acc.}&{0.722}&{0.758}&{0.686}&{0.765}&{0.750}&{0.738}&{0.794}&{0.706}&{0.737}&{0.747}&{0.713}&{\textbf{0.805}}\\

&{Prec.}&{0.726}&{0.753}&{0.704}&{0.786}&{0.782}&{0.741}&{0.810}&{0.698}&{0.721}&{0.744}&{0.704}&{\textbf{0.833}}\\

&{Rec.}&{0.900}&{\textbf{0.917}}&{0.833}&{0.734}&{0.838}&{0.795}&{0.880}&{0.706}&{0.723}&{0.717}&{0.683}&{0.863}\\

&{$F_1$}&{0.804}&{0.827}&{0.763}&{0.759}&{0.809}&{0.767}&{0.843}&{0.697}&{0.722}&{0.723}&{0.684}&{\textbf{0.848}}\\
\bottomrule
\end{tabular}
}
\end{table*}

We compare with the following baselines: 
\begin{itemize}
\item \textbf{LIWC}~\cite{pennebaker2015development} is a widely used text analysis method to extract lexicons falling into psycho-linguistic 
categories. The features are fed into different machine learning algorithms (Decision Tree, Logistic Regression, Random Forest and Naive Bayes) to achieve the best classification performance. 
\item \textbf{TextCNN}~\cite{DBLP:conf/emnlp/Kim14} utilizes convolutional neural networks to classify the posts.
\item \textbf{LSTM}~\cite{DBLP:conf/ijcai/MaGMKJWC16} 
capture the variation of contextual information of relevant posts over time for identifying rumors. 
\item \textbf{HAN}~\cite{DBLP:conf/naacl/YangYDHSH16} utilizes a hierarchical attention NN-framework, which encodes the post with word-level attention on each sentence and sentence-level attention on each document.
\item \textbf{BERT}~\cite{devlin2018bert} is a pre-trained language model based on deep bidirectional transformers, and it can be used to get the representation of the post text.
\item \textbf{Transformer}~\cite{vaswani2017attention} uses the self-attention mechanism to extract text context information and uses position encoding to determine the position of word vectors in the text. Transformer is a sequence to sequence model, which is not entirely suitable for our task, we thus only use its encoder here.
\item \textbf{EANN}~\cite{DBLP:conf/kdd/WangMJYXJSG18} is a GAN-based model that can extract event-invariant features and thus benefit the detection of newly arrived events. Note that different from the original EANN, we don't use pictures due to the lack of pictures in our dataset.
\item \textbf{RumorGAN}~\cite{DBLP:conf/www/Ma0W19} generates uncertain or conflicting voices, aiming to enhance the discriminator to learn stronger rumor representations from the augmented examples.
\item \textbf{dEFEND}~\cite{DBLP:conf/kdd/ShuCW0L19} adopts a deep hierarchical co-attention network to learn feature representations for both post contents and user comments.
\item \textbf{RVNN}~\cite{DBLP:conf/acl/WongGM18} learns discriminative features from contents by following their non-sequential propagation structure. RVNN includes a bottom-up and a top-down tree-structured network, denoted as RVNN$_{BU}$ and RVNN$_{TD}$ respectively.
\end{itemize}

LIWC is a traditional feature engineering method. 
TextCNN, LSTM, BERT, HAN, Transformer only uses the post contents, while dEFEND and RVNN exploit both the post contents and user comments. The learned representations then go through the fully-connected layer and softmax layer for classification.
EANN and RumorGAN use generative adversarial ideas to generate augmented examples to train a stronger detector. 



\subsection{Detection Performance}


We use the accuracy, precision, recall, and F1-score as evaluation metrics, and randomly choose 75\% of the instances for training and the remaining 25\% for testing.

Table 2 demonstrates the performance of all the compared models based on two datasets. The results show QSAN outperforms all the baselines, which confirms the advantage of the complex network and signed attention mechanism. On the CED dataset, deep learning-based methods outperform the traditional feature extraction methods. The methods incorporating comments, i.e., dEFEND, RVNN, and our QSAN, are generally better than the other solutions, indicating that comments are beneficial for detection. With the same input information, HAN, BERT, Transformer outperform generally their counterparts TextCNN and LSTM, and QSAN and dEFEND outperform RVNN, indicating the effectiveness of the attention mechanism. 


Similar trends are observed on FakeNews dataset, and the attention mechanism is even more effective. HAN, Bert, Transformer, dEFEND, and QSAN are generally better than the other competitors. 



\subsection{Ablation Test}
Here we evaluate the effectiveness of QSAN network components in improving the performance. To this end, several simplified variations of QSAN are implemented by replacing certain components:




\begin{itemize}
\item \textbf{QSAN-R-Co}: uses the real-valued word embeddings as the inputs, and replaces signed attention with the traditional real-valued co-attention.
\item \textbf{QSAN-R}: uses the real-valued word embeddings as inputs, with the remaining structure unchanged.
\item \textbf{QSAN-Co}: adopts a complex-valued co-attention module instead of the signed attention module of QSAN.

\end{itemize}

The results of the ablation test are reported in Figure 3. When comparing co-attention with signed-attention, we can observe that signed-attention achieves a consistent improvement over co-attention in both real and complex embeddings. On the CED dataset, the performance improvements are 0.7\% and 0.8\% in terms of $F_1$ and accuracy metrics respectively in real embeddings, and are 1.1\% and 1.2\% in complex embeddings. We can see similar trends on the FNews. It indicates the effectiveness of the signed-attention in capturing the comment semantics from different perspectives. Compared to the real-valued embeddings, the complex embeddings can provide better performance as they can better model the texts. E.g., on the CED dataset, the performance improves are 0.9\% and 1.0\% in terms of $F_1$ and accuracy with co-attention, and are 1.3\% and 1.4\% with signed-attention. On FNews, the trends are similar. QSAN outperforms QSAN-R, QSAN-Co, demonstrating that both the complex network and the signed attention mechanism are able to contribute to the superior performance of QSAN.

\begin{figure}
\vspace{-0.4cm}
\setlength{\abovedisplayshortskip}{0pt}
\begin{center}
\label{fig:parameter}
\begin{tabular}{c c}
\subfigure{\includegraphics[width=0.47\linewidth]{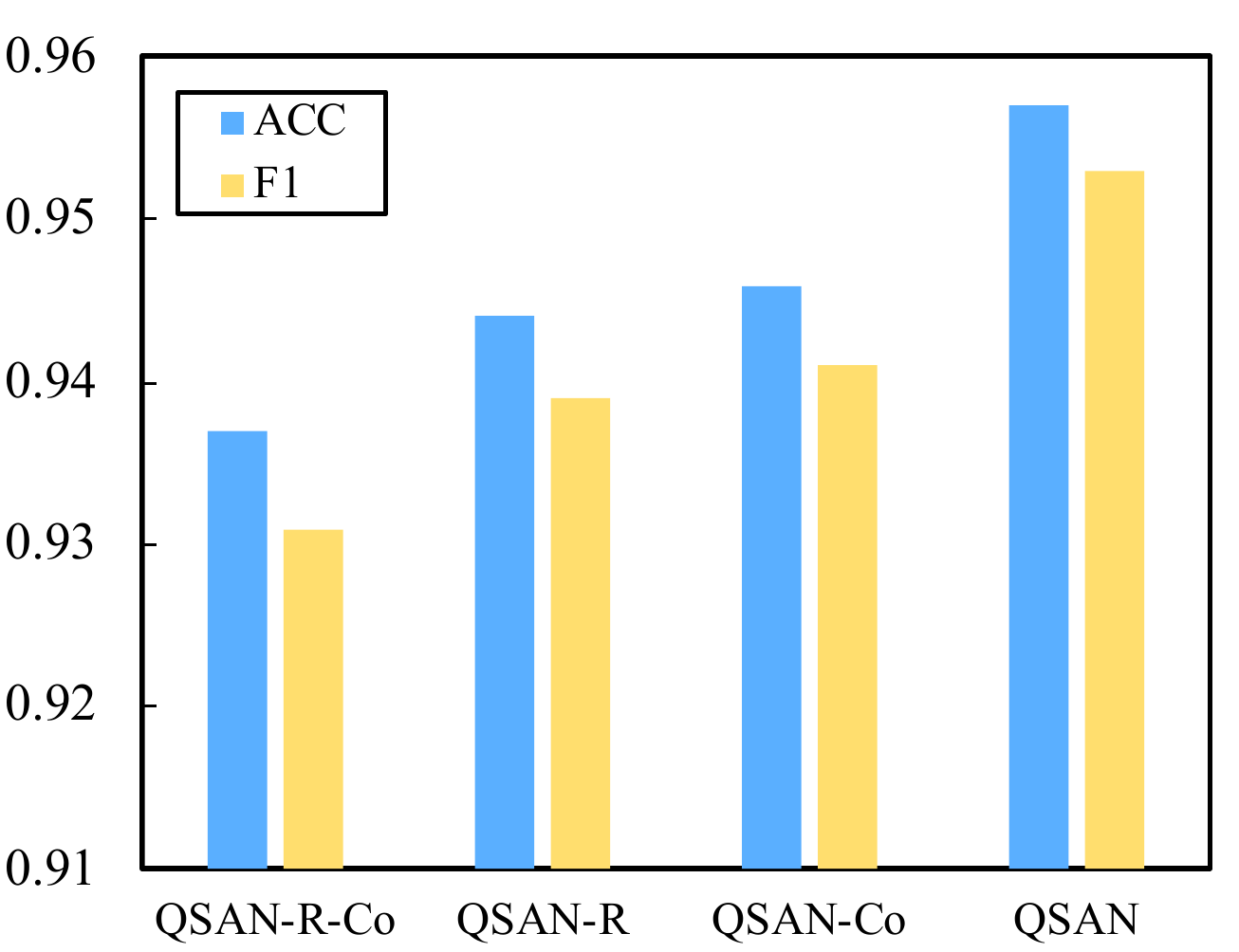}} & \subfigure{\includegraphics[width=0.47\linewidth]{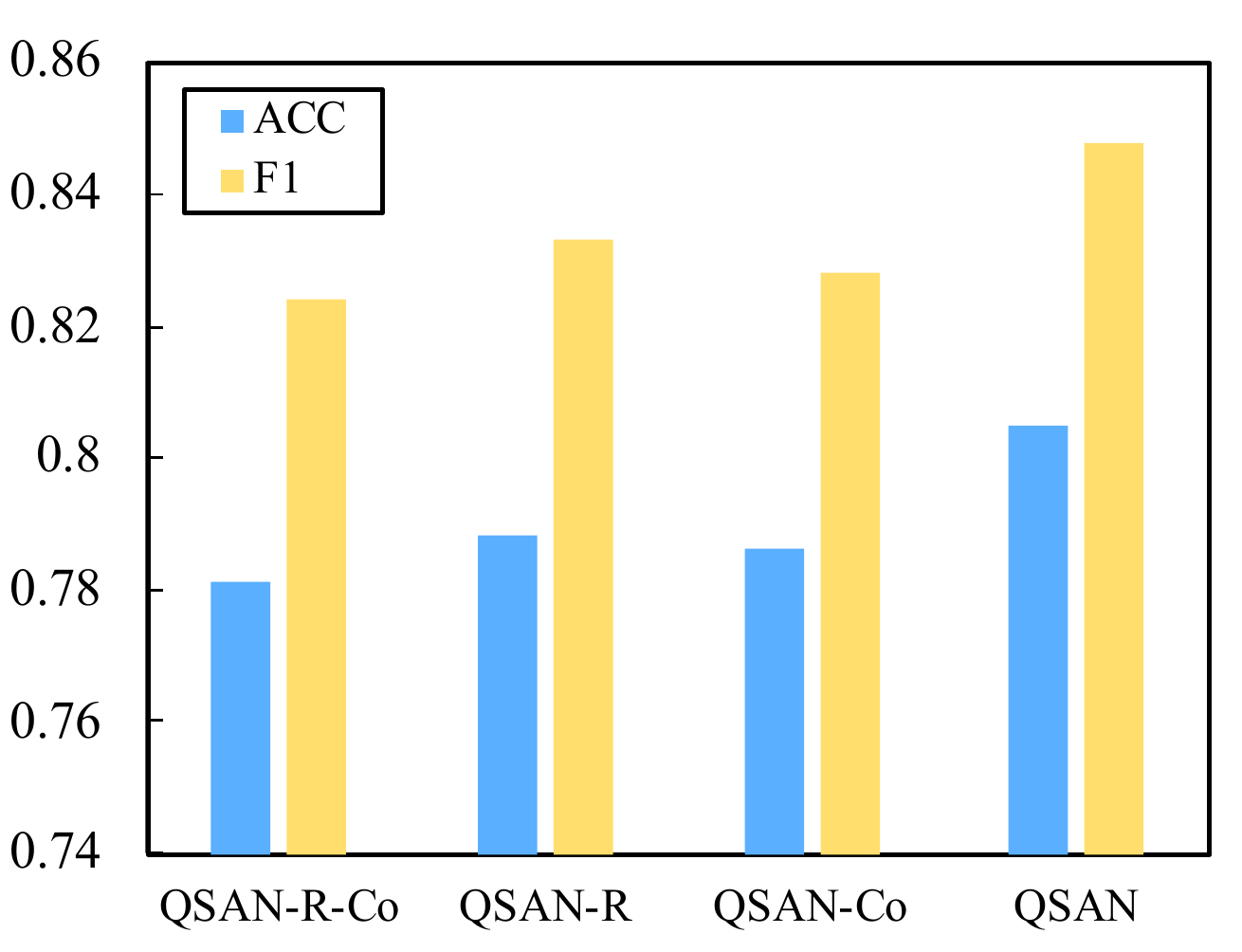}} \\
(a) CED & (b) FNews\\
\end{tabular}
\caption{Impact analysis of the signed attention and complex network for false information detection.}
\end{center}
\vspace{-0.5cm}
\end{figure}

\subsection{Post-Comment Relationship Analysis}
We evaluate the effectiveness and explainability of the signed attention from two perspectives: (1) whether the captured comments can show "supporting" and "opposing" views; (2) whether the "important" comments can be distinguished from those "unimportant". We first illustrate how QSAN captures post-comment relationships and provide meaningful comments in this subsection, and evaluate the meaningful comments with human judgment in the next. 

With the help of QSAN, we can investigate different relationships between the comments and the post, which is enabled by the attention weights in Eq. (11) and Eq. (12). To better understand how to use such attention weights, we start from the unnormalized weights before the ``+softmax" and ``-softmax" functions. The reason why we look at the unnormalized weights is that we need to retain the plus and minus signs of the weights to show they indicate positive and negative correlations between comments and posts respectively. For the sake of simplicity, the unnormalized weights of comments are represented as $({re}^+, {im}^+ |{re}^-, {im}^-)$, where $({re}^+, {im}^+)$ and $({re}^-, {im}^-)$ denote the real and imaginary parts of the unnormalized weights. That is, $({re}^+, {im}^+ |{re}^-, {im}^-)$ denotes the real and imaginary parts of $(\bm{W}^{c_{pos}}(\bm{H}^{c}) | \bm{W}^{c_{neg}}(\bm{H}^{c}))$ presented in Eq. (11) and Eq. (12).

For ease of interpretation, we give a real example from our training instances in Figure 4 to illustrate the meanings of the unnormalized weights. It can be observed that the comments are categorized into three types: supporting (in the green text box), opposing (in the red text box), and neutral (in the yellow text box), whose weights have the following characteristics.  
\begin{figure}
\vspace{-0.4cm}
\begin{center}
\includegraphics[width=0.45\textwidth]{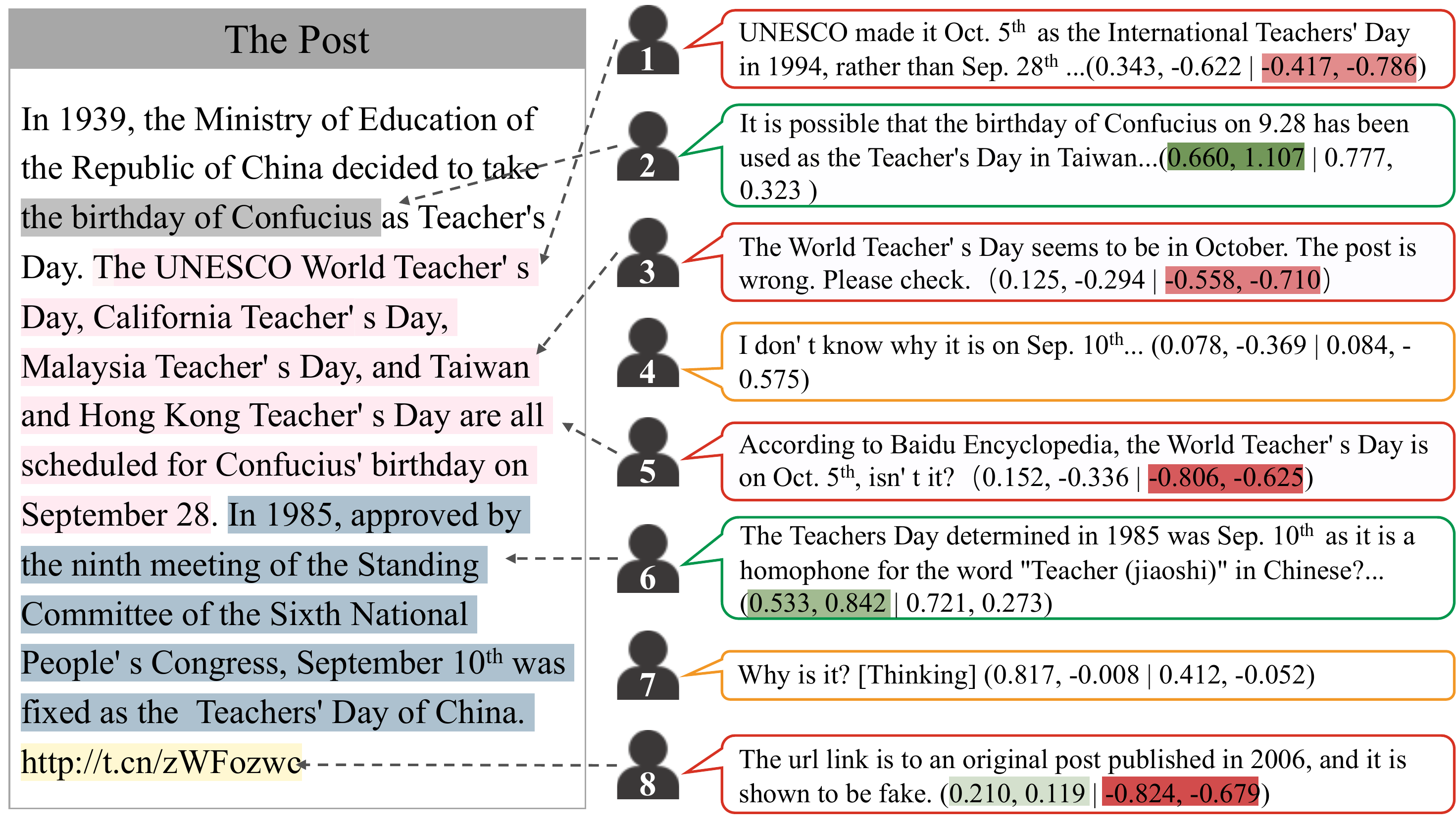}
\caption{The user comments are related to specific sentences in the post, and red, green and yellow text boxes contain opposing, supporting and neutral comments respectively. The darker the color of unnormalized weights is, the more important the comment is in its corresponding view (supporting or opposing).} \label{fig1}
\end{center}
\vspace{-0.6cm}
\end{figure}

(1) For the supporting comment (e.g., Comment 2 and 6), it typically has positive values for both ${re}^+$ and ${im}^+$, indicating the comment is positively correlated to the post content. 
 
(2) For the opposing comment (e.g., Comment 1, 3, 5), it typically has negative values for both ${re}^-$ and ${im}^-$, indicating the comment is negatively correlated to the post content. 

(3) For a comment that satisfies both condition (1) and (2) (e.g., Comment 8), to decide whether it is supporting or opposing, we would further account for its importance in two views. The supporting importance and opposing importance are defined as the modulus $s^+=\sqrt{{({re}^+)}^2+{({im}^+)}^2}$ and $s^-=\sqrt{{({re}^-)}^2+{({im}^-)}^2}$ respectively. We then rank $s^+$ of a specific comment with those of the other comments in the positive view, and rank $s^-$ of a specific comment with those of all the other comments in the negative view. A larger modulus indicates a higher rank. If $s^-$ ranked higher than $s^+$, it indicates the negative correlation dominates the comment, and thus it is opposing, such as Comment 8.

(4) For the remaining cases that there are both positive and negative values for ${re}^+$ and ${im}^+$ and for ${re}^-$ and ${im}^-$ (e.g., Comment 4, 7), the comment is generally neutral.

The above real example confirms the motivation of the signed attention described in Sec. 5.3, and gives hints on how to determine the stance of a comment. The comment with a larger $s^+$ (resp. $s^-$ ) indicates that it is more likely to be a supporting (resp. opposing) one than those with smaller $s^+$ (resp.  $s^-$). We will evaluate this argument with human judgment in the next subsection.

There are both supporting and opposing importance for each comment, which can be understood as a comment may present multiple views at the same time. It would be interesting to evaluate the overall importance of a comment by comprehensively considering its importance in both views. However, $s^-$ and $s^+$ are obtained based on unnormalized weights and thus cannot be directly compared. To address this issue, we use their normalized weights ${a}^{c_{pos}}$ and ${a}^{c_{neg}}$ introduced in Eq. (11) and Eq. (12) instead. Then the amplitude of the positive attentions is obtained through $s_n^+ = \sqrt{{{re}({a}^{c_{pos}})}^2+{{im({a}^{c_{pos}})}^2}}$, while the amplitude of the negative attentions is obtained through $s_n^- = \sqrt{{{re}({a}^{c_{neg}})}^2+{{im({a}^{c_{neg}})}^2}}$.

Given $s_n^+$ and $s_n^-$ of a comment, we propose to use $imp = \lvert {s_n^+}-{s_n^-}\rvert $ for measuring the overall importance of this comment. The larger the value is, the more important the comment is. The intuition behind is that if a comment has very close $s_n^+$ and $s_n^-$, it may be ambiguous and not able to provide very discriminative features. On the contrary, if $s_n^+$ distincts from $s_n^-$, it tends to provide a very clear voice for false information detection. 
We will evaluate the importance of comments in the following subsection.

\subsection{Exlainability Evaluation}

Inspired by the explainability evaluation methods in~\cite{DBLP:conf/kdd/ShuCW0L19}, we conduct more comprehensive and detailed list-wise and item-wise human evaluations. In addition, as dEFEND~\cite{DBLP:conf/kdd/ShuCW0L19} is also able to extract the important comments for explanations, we will compare the characteristics of the important comments captured by QSAN and dEFEND, which would demonstrate the differences between these two models. For human evaluations, we recruited 3 undergraduates who majored in Computer Science and Mathematics as annotators. As their native language is Chinese, we ask them to annotate a total of 100 pieces of false information with comments in the CED dataset. The informative comments captured by QSAN and dEFEND will be evaluated by all annotators with a three-person voting approach.


\subsubsection{List-wise Evaluation}
\begin{figure}

\setlength{\abovedisplayshortskip}{0pt}
\begin{center}
\label{fig:parameter}
\begin{tabular}{c c}
\subfigure{\includegraphics[width=0.48\linewidth]{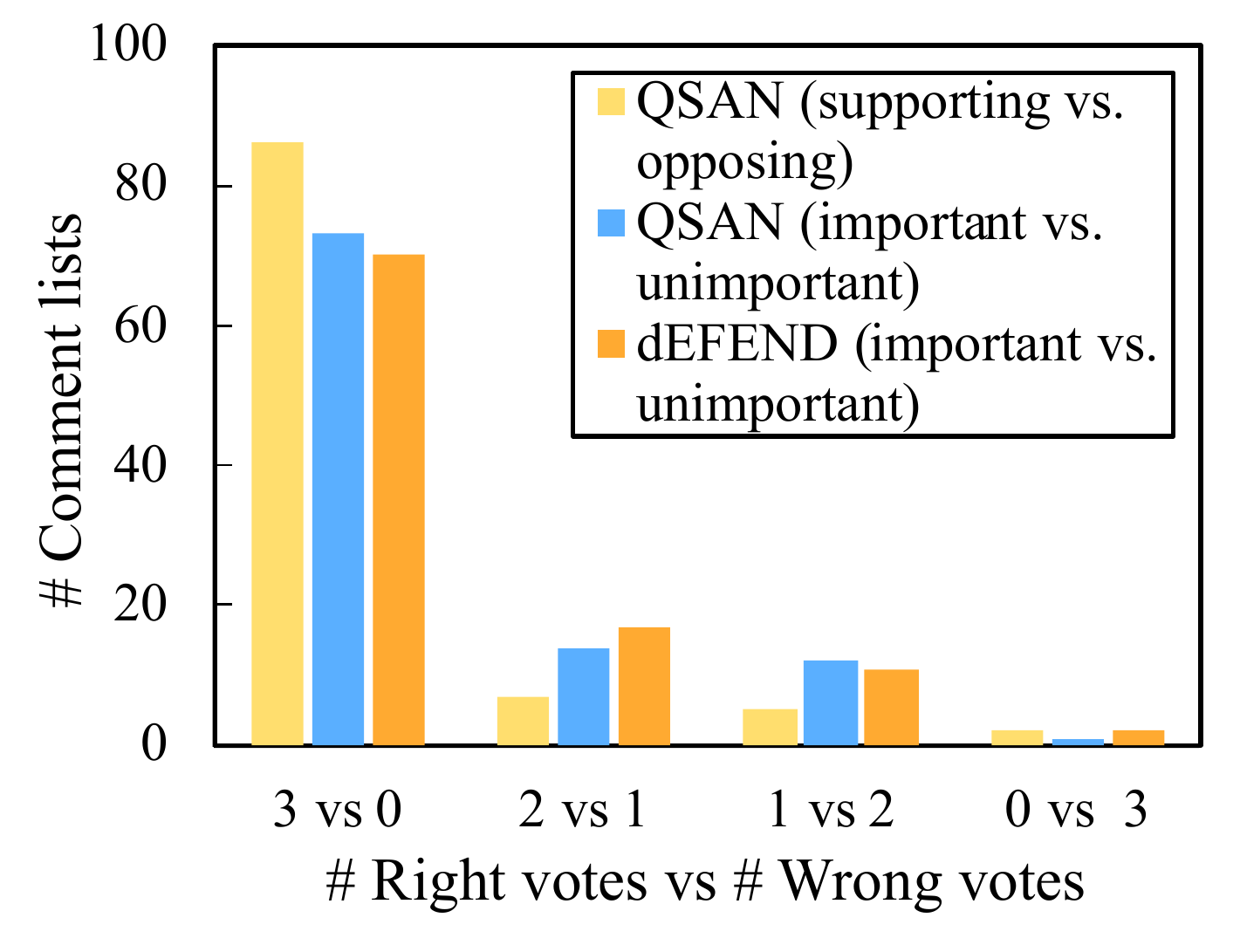}} & \subfigure{\includegraphics[width=0.48\linewidth]{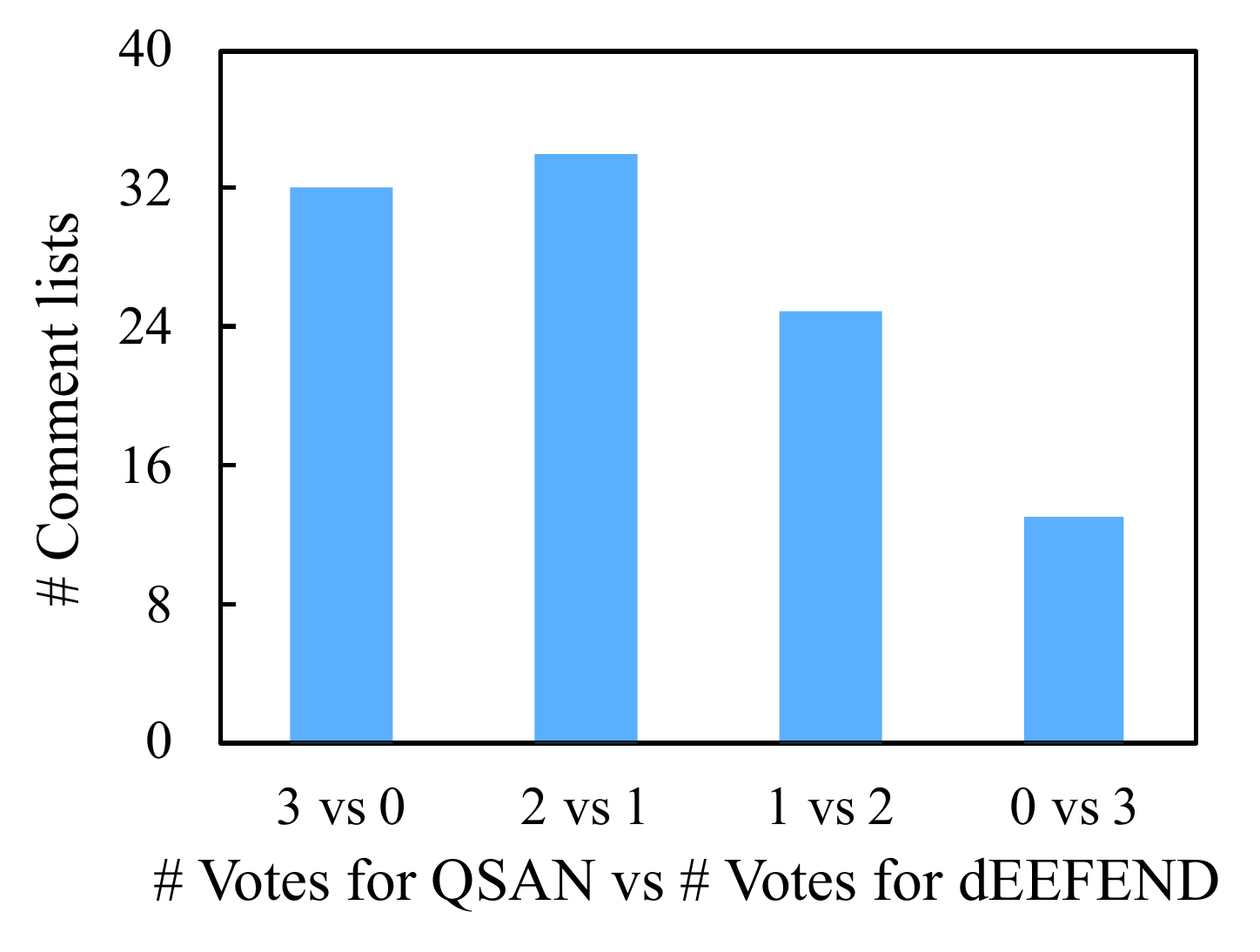}} \\
(a)& (b)\\
\end{tabular}
\vspace{-0.4cm}
\caption{The explainability evaluation results for list-wise comments. Figure 5(a) shows the winning count of voting for distinguishing important lists from unimportant lists, and supporting lists from opposing lists. Figure 5(b) indicates the voting ratio to show who gets better important lists, QSAN or dEFEND.}
\end{center}
\vspace{-0.4cm}
\end{figure}

\begin{figure}
\setlength{\abovedisplayshortskip}{0pt}
\begin{center}
\label{fig:parameter}
\begin{tabular}{c c}
\subfigure{\includegraphics[width=0.48\linewidth]{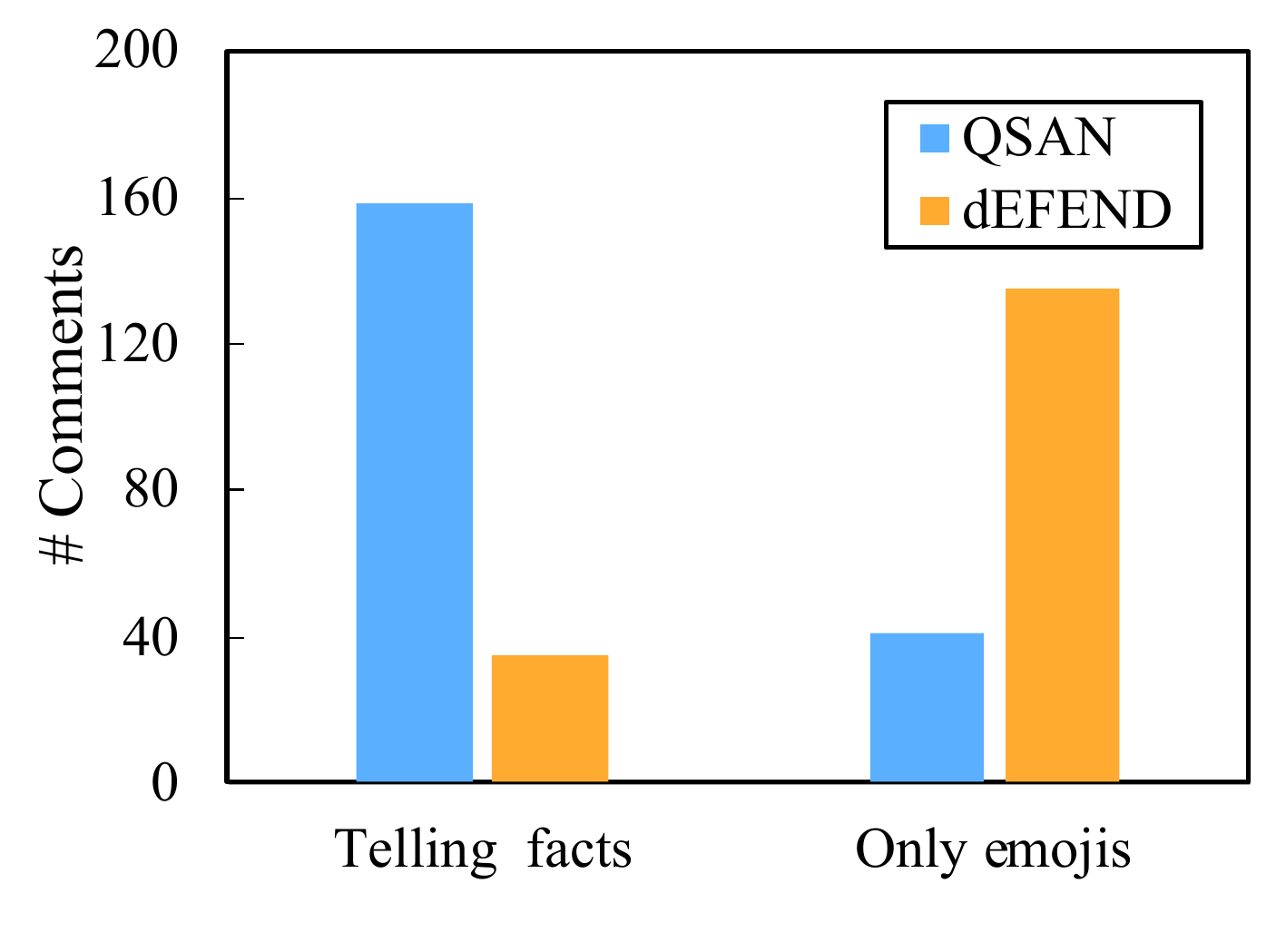}} & \subfigure{\includegraphics[width=0.48\linewidth]{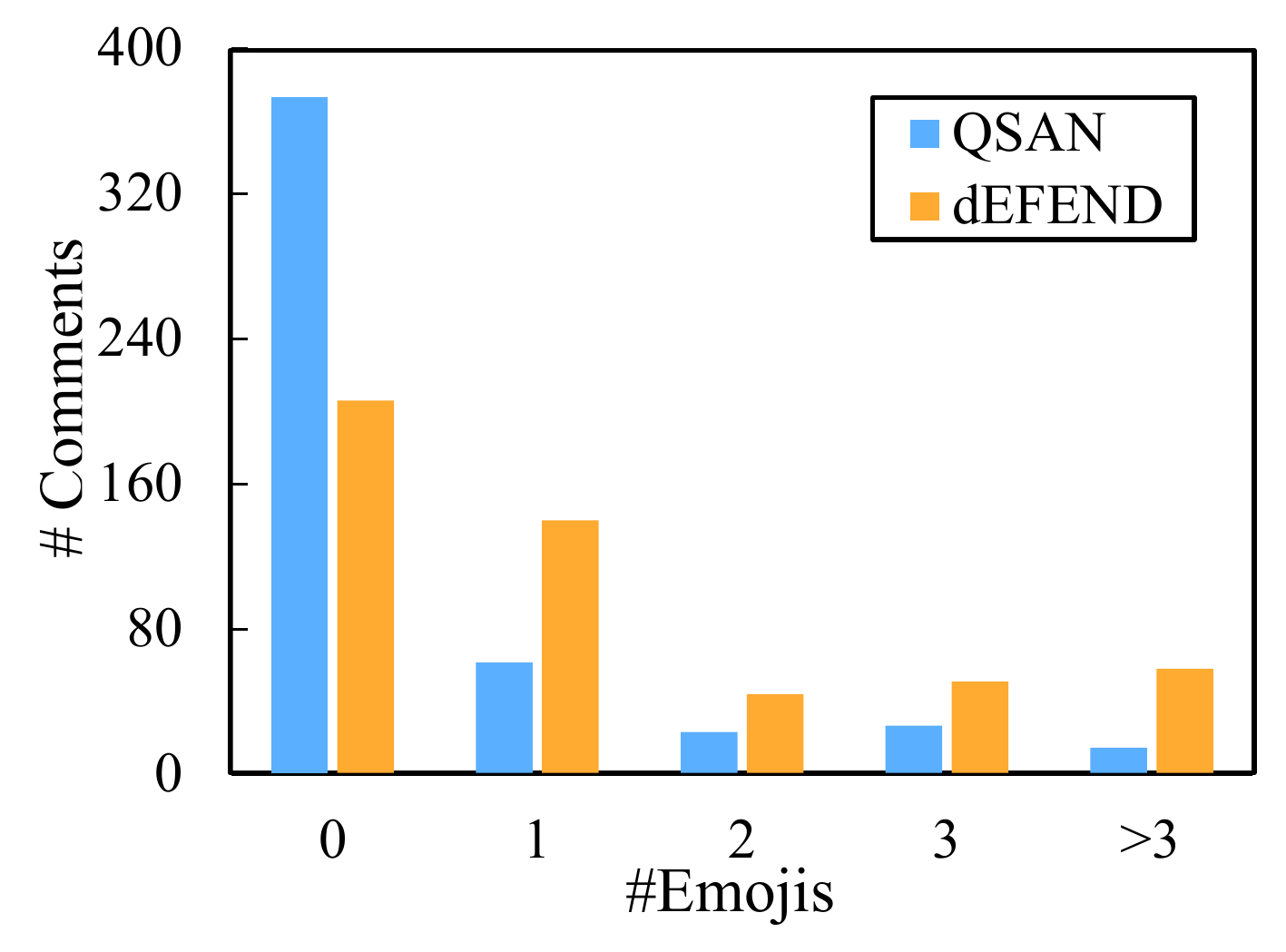}} \\
(a)& 
(b)
\\
\end{tabular}
\vspace{-0.4cm}
\caption{The comparison of top-5 important comments produced by QSAN and dEFEND. (a) Counts of comments in two different groups categorized by human judgment. (b) Counts of important comments with different numbers of emojis.}
\end{center}
\vspace{-0.4cm}
\end{figure}

To demonstrate the coarse-grained explainability performance of the model, we conduct a list-wise evaluation from the following two perspectives.

\textbf{Evaluation for "Supporting" or "Opposing".} For each of the 100 pieces of information, we conduct an explainability experiment of distinguishing "supporting" or "opposing" comment lists. Specifically, for the comments corresponding to a post, we calculate their $s_n^+$ and $s_n^-$ according to the methods introduced in Sec. 6.5, and sort the comments with their $s_n^+$ and $s_n^-$ values respectively. Then, the comments with the top-$k$ largest $s_n^+$ constitute the "supporting" comment list $(list_{QP})$, and those with the top-$k$ largest $s_n^-$ constitute the "opposing" comment list $(list_{QN})$ (we set $k=5$ in the experiment section). We randomly shuffle the order of $list_{QP}$ and $list_{QN}$ to remove the position bias, and present them to all three annotators to decide which one is supporting. The voting results are shown in Figure 5(a), we can observe that for 86 out of 100 posts, the supporting comment lists are correctly labeled by all three voters (3 vs. 0), and only 2 comment lists get three wrong votes (0 vs. 3), demonstrating the effectiveness of the signed attention mechanism in capturing the supporting and opposing relationships.

\textbf{Evaluation for "Important" or "Unimportant". } Here we deploy two human evaluation tasks: (1) whether the important comment list and unimportant comment list can be distinguished; (2) given two important comment lists captured by QSAN and dEFEND respectively, which one is better, i.e., more important?

For the first task, the important and unimportant comment lists captured by QSAN are obtained by ranking $imp$ introduced in Sec. 6.5. Specifically, comments with the top-$k$ largest (resp. smallest) $imp$ constitute the important (resp. unimportant) comment list. According to~\cite{DBLP:conf/kdd/ShuCW0L19}, the two comment lists captured by dEFEND are obtained according to attention weights. For each pair of the important and unimportant comment lists, we randomly shuffle their order and present them to all three annotators to vote. Figure 5(a) shows that QSAN has comparable or better performance than dEFEND in this task. In the case of all 3 voters voting correctly (3 vs. 0) and the case of no voter voting correctly (0 vs. 3), QSAN performers slightly better than dEFEND (i.e., 73>70 and 1<2 respectively). And in the third case of 2 voters voting correctly (2 vs. 1), dEFEND shows a bit better performance (i.e., 17>14). 

For the second task, we mix and shuffle the order of the two "important" comment lists selected by QSAN and dEFEND, and the voting results are shown in Figure 5(b). We can observe that QSAN outperforms dEFEND in 66 out of 100 pairs of important comment lists (32 for 3 vs. 0 and 34 for 2 vs. 1), indicating QSAN is better at capturing the important comments. 

Although both QSAN and dEFEND can produce important comments, the characteristics of important comments they choose are different. We 
find that dEFEND are more likely to choose comments with emojis while QSAN tends to choose comments with facts. We thus label and count these two types of comments captured by QSAN and dEFEND, and the results are shown in Figure 6. Note that here we omit the important comments that don't fall into these two categories. We have the following observations. (1) According to Figure 6(a), QSAN is more likely to choose comments which tell facts than dEFEND (i.e., 158>35), while dEFEND selects a larger number of comments that contain nothing but emojis (i.e., 41<135). (2) Figure 6(b) shows the distribution of important comments with a varied number of emojis. And it shows that dEFEND is more likely to regard emojis as important clues to distinguish true and false posts than QSAN (i.e., for one emoji: 140>62, for two: 44>23, for three: 51>27, for more than 3: 59>14).


To summarize, compared to dEFEND, the important comments produced by QSAN can tell more facts for the task, and thus may be better in line with human judgments. It implies that QSAN is good at capturing the complex semantics and relationships from the post and comments due to the complex networks and signed attention. This may also explain why QSAN receives more votes in Figure 5(b) for getting better important comment lists. 

\subsubsection{Item-wise Evaluation}
In addition to list-wise evaluations, we also conduct fine-grained item-wise experiments. 

\textbf{"Supporting" or "Opposing" Evaluation Setting.} For each piece of information, we mix and shuffle the order of the captured top 5 supporting comments and top 5 opposing comments. Then a list of "mixed" 10 comments is presented to all three annotators. Each annotator is required to assign a score from (-2, -1, 0, 1, 2) to each comment, where 0 means "strong opposing", -1 means "opposing", 0 means "neutral", 1 means "supporting", and 2 means "strong supporting". 
We then sum the scores for each comment. The sum is positive (resp. negative) indicates the comment is supporting (resp. opposing). If the sum is equal to 0, we ask a fourth person to make a judgment. Finally, all the comments are labeled as supporting or opposing, and we use the labels as ground truth. 


\begin{table}
\centering
\caption{Explainability Evaluation for Item-wise Comments}
\scriptsize
\setlength{\tabcolsep}{4mm}{
\begin{tabular}{ c |c c c c c }
\toprule 
\textbf{Method} & Class & Acc. & Prec. & Rec. & $F_1$\cr
\midrule
\multirow{2}*{\textbf{QSAN}}&{Supporting}&{\multirow{2}*{0.727}}&{0.704}&{0.784}&{0.742}\\
&{Opposing}& &{0.704}&{0.784}&{0.742}\\
\midrule
\midrule
\multirow{2}*{\textbf{QSAN}}&{Important}&{\multirow{2}*{\textbf{0.712}}}&{\textbf{0.729}}&{\textbf{0.674}}&{\textbf{0.701}}\\
&{Unimportant}& &{\textbf{0.697}}&{0.750}&{\textbf{0.723}}\\
\midrule
\multirow{2}*{\textbf{dEFEND}}&{Important}&{\multirow{2}*{0.692}}&{0.726}&{0.612}&{0.667}\\
&{Unimportant}& &{0.667}&{\textbf{0.768}}&{0.714}\\
\bottomrule
\end{tabular}}
\vspace{-0.4cm}
\end{table}




\textbf{"Important" or "Unimportant" Evaluation Setting.} Here we mix and shuffle the comments from the important and unimportant comment lists corresponding to a post, and ask each annotator to assign a score from (-2, -1, 0, 1, 2) to a comment, which means "not helpful", "not much helpful", "not sure", "a bit helpful", "helpful" respectively. We also sum the scores for each comment, and decide whether it is helpful or not according to the signs of the sum. If the sum equals to 0, a fourth person is asked for judging. We compare the captured comments by QSAN and dEFEND with the ground truth labels. The results are shown in Table 3. It can be observed that: (1) QSAN can distinguish between "supporting" and "opposing" comments with an accuracy of 0.727, which validates the effectiveness of QSAN in capturing the stances. (2) In terms of importance distinction, QSAN outperforms dEFEND in almost all the metrics except the recall of "unimportant". It confirms that QSAN can better capture the important comments than dEFEND.

\section{Conclusion}

In this work, we propose a Quantum-probability based Signed Attention Network (QSAN) to capture the complex semantic relationships between the post and the comments. In particular, a novel signed attention mechanism is devised to take into account both the importance and the stance of comments to improve the detection performance. In addition, QSAN can provide the benefits of explainability from two aspects, model transparency and informative comments as explanations. Extensive experiments on English and Chinese datasets demonstrate the effectiveness and explainability of QSAN. For future directions, it would be promising to introduce the stance labels of comments and user credibility information to further improve the performance of false information detection.



\section*{Acknowledgments}

This work was supported by the National Key R\&D Program of China (No.2017YFB0803301)
and Natural Science Foundation of China (No.61976026,
No.U1836215) and 111 Project (B18008).

\bibliographystyle{ACM-Reference-Format}
\bibliography{sample-base}


\begin{thebibliography}{51}


\ifx \showCODEN    \undefined \def \showCODEN     #1{\unskip}     \fi
\ifx \showDOI      \undefined \def \showDOI       #1{#1}\fi
\ifx \showISBNx    \undefined \def \showISBNx     #1{\unskip}     \fi
\ifx \showISBNxiii \undefined \def \showISBNxiii  #1{\unskip}     \fi
\ifx \showISSN     \undefined \def \showISSN      #1{\unskip}     \fi
\ifx \showLCCN     \undefined \def \showLCCN      #1{\unskip}     \fi
\ifx \shownote     \undefined \def \shownote      #1{#1}          \fi
\ifx \showarticletitle \undefined \def \showarticletitle #1{#1}   \fi
\ifx \showURL      \undefined \def \showURL       {\relax}        \fi
\providecommand\bibfield[2]{#2}
\providecommand\bibinfo[2]{#2}
\providecommand\natexlab[1]{#1}
\providecommand\showeprint[2][]{arXiv:#2}

\bibitem[\protect\citeauthoryear{Aerts and Sozzo}{Aerts and Sozzo}{2014}]%
        {aerts2014quantum}
\bibfield{author}{\bibinfo{person}{Diederik Aerts} {and}
  \bibinfo{person}{Sandro Sozzo}.} \bibinfo{year}{2014}\natexlab{}.
\newblock \showarticletitle{Quantum entanglement in concept combinations}.
\newblock \bibinfo{journal}{\emph{International Journal of Theoretical
  Physics}} \bibinfo{volume}{53}, \bibinfo{number}{10} (\bibinfo{year}{2014}).
\newblock


\bibitem[\protect\citeauthoryear{Al~Janaby and Abed}{Al~Janaby and
  Abed}{2011}]%
        {al2011syntactic}
\bibfield{author}{\bibinfo{person}{Hameed~M Al~Janaby} {and}
  \bibinfo{person}{Ammar~A Abed}.} \bibinfo{year}{2011}\natexlab{}.
\newblock \showarticletitle{Syntactic Ambiguity in Newspaper Headlines}.
\newblock \bibinfo{journal}{\emph{Journal of the Faculty of Collective
  Knowledge}} (\bibinfo{year}{2011}).
\newblock


\bibitem[\protect\citeauthoryear{Antova, Koch, and Olteanu}{Antova
  et~al\mbox{.}}{2009}]%
        {antova200910}
\bibfield{author}{\bibinfo{person}{Lyublena Antova}, \bibinfo{person}{Christoph
  Koch}, {and} \bibinfo{person}{Dan Olteanu}.} \bibinfo{year}{2009}\natexlab{}.
\newblock \showarticletitle{${10^{(10^{6})}}$ worlds and beyond: efficient
  representation and processing of incomplete information.}
\newblock \bibinfo{journal}{\emph{The VLDB Journal}} \bibinfo{volume}{18},
  \bibinfo{number}{5} (\bibinfo{year}{2009}), \bibinfo{pages}{1021}.
\newblock


\bibitem[\protect\citeauthoryear{Bahdanau, Cho, and Bengio}{Bahdanau
  et~al\mbox{.}}{2015}]%
        {bahdanau2014neural}
\bibfield{author}{\bibinfo{person}{Dzmitry Bahdanau},
  \bibinfo{person}{Kyunghyun Cho}, {and} \bibinfo{person}{Yoshua Bengio}.}
  \bibinfo{year}{2015}\natexlab{}.
\newblock \showarticletitle{Neural machine translation by jointly learning to
  align and translate}. In \bibinfo{booktitle}{\emph{ICLR}}.
\newblock


\bibitem[\protect\citeauthoryear{Bollen, Mao, and Zeng}{Bollen
  et~al\mbox{.}}{2011}]%
        {bollen2011twitter}
\bibfield{author}{\bibinfo{person}{Johan Bollen}, \bibinfo{person}{Huina Mao},
  {and} \bibinfo{person}{Xiaojun Zeng}.} \bibinfo{year}{2011}\natexlab{}.
\newblock \showarticletitle{Twitter mood predicts the stock market}.
\newblock \bibinfo{journal}{\emph{Journal of computational science}}
  \bibinfo{volume}{2}, \bibinfo{number}{1} (\bibinfo{year}{2011}),
  \bibinfo{pages}{1--8}.
\newblock


\bibitem[\protect\citeauthoryear{Bruza, Kitto, and McEvoy}{Bruza
  et~al\mbox{.}}{2008}]%
        {bruza2008entangling}
\bibfield{author}{\bibinfo{person}{Peter Bruza}, \bibinfo{person}{Kirsty
  Kitto}, {and} \bibinfo{person}{Doug McEvoy}.}
  \bibinfo{year}{2008}\natexlab{}.
\newblock \showarticletitle{Entangling words and meaning}. In
  \bibinfo{booktitle}{\emph{Quantum Interaction: Proceedings of the Second
  Quantum Interaction Symposium (QI-2008):}}. College Publications,
  \bibinfo{pages}{118--124}.
\newblock


\bibitem[\protect\citeauthoryear{Bruza, Kitto, Nelson, and McEvoy}{Bruza
  et~al\mbox{.}}{2009}]%
        {bruza2009there}
\bibfield{author}{\bibinfo{person}{Peter Bruza}, \bibinfo{person}{Kirsty
  Kitto}, \bibinfo{person}{Douglas Nelson}, {and} \bibinfo{person}{Cathy
  McEvoy}.} \bibinfo{year}{2009}\natexlab{}.
\newblock \showarticletitle{Is there something quantum-like about the human
  mental lexicon?}
\newblock \bibinfo{journal}{\emph{Journal of Mathematical Psychology}}
  \bibinfo{volume}{53}, \bibinfo{number}{5} (\bibinfo{year}{2009}),
  \bibinfo{pages}{362--377}.
\newblock


\bibitem[\protect\citeauthoryear{Chaudhari, Polatkan, Ramanath, and
  Mithal}{Chaudhari et~al\mbox{.}}{2019}]%
        {chaudhari2019attentive}
\bibfield{author}{\bibinfo{person}{Sneha Chaudhari}, \bibinfo{person}{Gungor
  Polatkan}, \bibinfo{person}{Rohan Ramanath}, {and} \bibinfo{person}{Varun
  Mithal}.} \bibinfo{year}{2019}\natexlab{}.
\newblock \showarticletitle{An attentive survey of attention models}.
\newblock \bibinfo{journal}{\emph{arXiv preprint arXiv:1904.02874}}
  (\bibinfo{year}{2019}).
\newblock


\bibitem[\protect\citeauthoryear{Dalvi, R{\'e}, and Suciu}{Dalvi
  et~al\mbox{.}}{2009}]%
        {dalvi2009probabilistic}
\bibfield{author}{\bibinfo{person}{Nilesh Dalvi}, \bibinfo{person}{Christopher
  R{\'e}}, {and} \bibinfo{person}{Dan Suciu}.} \bibinfo{year}{2009}\natexlab{}.
\newblock \showarticletitle{Probabilistic databases: diamonds in the dirt}.
\newblock \bibinfo{journal}{\emph{Commun. ACM}} \bibinfo{volume}{52},
  \bibinfo{number}{7} (\bibinfo{year}{2009}), \bibinfo{pages}{86--94}.
\newblock


\bibitem[\protect\citeauthoryear{Devlin, Chang, Lee, and Toutanova}{Devlin
  et~al\mbox{.}}{2019}]%
        {devlin2018bert}
\bibfield{author}{\bibinfo{person}{Jacob Devlin}, \bibinfo{person}{Ming-Wei
  Chang}, \bibinfo{person}{Kenton Lee}, {and} \bibinfo{person}{Kristina
  Toutanova}.} \bibinfo{year}{2019}\natexlab{}.
\newblock \showarticletitle{Bert: Pre-training of deep bidirectional
  transformers for language understanding}. In
  \bibinfo{booktitle}{\emph{NAACL-HLT}}. \bibinfo{pages}{4171--4186}.
\newblock


\bibitem[\protect\citeauthoryear{Du, Liu, and Hu}{Du et~al\mbox{.}}{2020}]%
        {DBLP:journals/cacm/DuLH20}
\bibfield{author}{\bibinfo{person}{Mengnan Du}, \bibinfo{person}{Ninghao Liu},
  {and} \bibinfo{person}{Xia Hu}.} \bibinfo{year}{2020}\natexlab{}.
\newblock \showarticletitle{Techniques for interpretable machine learning}.
\newblock \bibinfo{journal}{\emph{Commun. {ACM}}} \bibinfo{volume}{63},
  \bibinfo{number}{1} (\bibinfo{year}{2020}), \bibinfo{pages}{68--77}.
\newblock


\bibitem[\protect\citeauthoryear{Fisher, Cox, and Hermann}{Fisher
  et~al\mbox{.}}{2016}]%
        {fisher2016pizzagate}
\bibfield{author}{\bibinfo{person}{Marc Fisher}, \bibinfo{person}{John~Woodrow
  Cox}, {and} \bibinfo{person}{Peter Hermann}.}
  \bibinfo{year}{2016}\natexlab{}.
\newblock \showarticletitle{Pizzagate: From rumor, to hashtag, to gunfire in
  DC}.
\newblock \bibinfo{journal}{\emph{Washington Post}}  \bibinfo{volume}{6}
  (\bibinfo{year}{2016}).
\newblock


\bibitem[\protect\citeauthoryear{Goddard and Wierzbicka}{Goddard and
  Wierzbicka}{1994}]%
        {goddard1994semantic}
\bibfield{author}{\bibinfo{person}{Cliff Goddard} {and} \bibinfo{person}{Anna
  Wierzbicka}.} \bibinfo{year}{1994}\natexlab{}.
\newblock \bibinfo{booktitle}{\emph{Semantic and lexical universals: Theory and
  empirical findings}}. Vol.~\bibinfo{volume}{25}.
\newblock \bibinfo{publisher}{John Benjamins Publishing}.
\newblock


\bibitem[\protect\citeauthoryear{Grinberg, Joseph, Friedland, Swire-Thompson,
  and Lazer}{Grinberg et~al\mbox{.}}{2019}]%
        {grinberg2019fake}
\bibfield{author}{\bibinfo{person}{Nir Grinberg}, \bibinfo{person}{Kenneth
  Joseph}, \bibinfo{person}{Lisa Friedland}, \bibinfo{person}{Briony
  Swire-Thompson}, {and} \bibinfo{person}{David Lazer}.}
  \bibinfo{year}{2019}\natexlab{}.
\newblock \showarticletitle{Fake news on Twitter during the 2016 US
  presidential election}.
\newblock \bibinfo{journal}{\emph{Science}} \bibinfo{volume}{363},
  \bibinfo{number}{6425} (\bibinfo{year}{2019}), \bibinfo{pages}{374--378}.
\newblock


\bibitem[\protect\citeauthoryear{Guo, Cao, Zhang, Guo, and Li}{Guo
  et~al\mbox{.}}{2018}]%
        {guo2018rumor}
\bibfield{author}{\bibinfo{person}{Han Guo}, \bibinfo{person}{Juan Cao},
  \bibinfo{person}{Yazi Zhang}, \bibinfo{person}{Junbo Guo}, {and}
  \bibinfo{person}{Jintao Li}.} \bibinfo{year}{2018}\natexlab{}.
\newblock \showarticletitle{Rumor detection with hierarchical social attention
  network}. In \bibinfo{booktitle}{\emph{CIKM}}. \bibinfo{pages}{943--951}.
\newblock


\bibitem[\protect\citeauthoryear{Gupta, Lamba, Kumaraguru, and Joshi}{Gupta
  et~al\mbox{.}}{2013}]%
        {DBLP:conf/www/0003LKJ13}
\bibfield{author}{\bibinfo{person}{Aditi Gupta}, \bibinfo{person}{Hemank
  Lamba}, \bibinfo{person}{Ponnurangam Kumaraguru}, {and}
  \bibinfo{person}{Anupam Joshi}.} \bibinfo{year}{2013}\natexlab{}.
\newblock \showarticletitle{Faking Sandy: characterizing and identifying fake
  images on Twitter during Hurricane Sandy}. In
  \bibinfo{booktitle}{\emph{WWW}}. \bibinfo{pages}{729--736}.
\newblock


\bibitem[\protect\citeauthoryear{Jin, Cao, Zhang, and Luo}{Jin
  et~al\mbox{.}}{2016a}]%
        {DBLP:conf/aaai/JinCZL16}
\bibfield{author}{\bibinfo{person}{Zhiwei Jin}, \bibinfo{person}{Juan Cao},
  \bibinfo{person}{Yongdong Zhang}, {and} \bibinfo{person}{Jiebo Luo}.}
  \bibinfo{year}{2016}\natexlab{a}.
\newblock \showarticletitle{News Verification by Exploiting Conflicting Social
  Viewpoints in Microblogs}. In \bibinfo{booktitle}{\emph{AAAI}}.
\newblock


\bibitem[\protect\citeauthoryear{Jin, Cao, Zhang, Zhou, and Tian}{Jin
  et~al\mbox{.}}{2016b}]%
        {jin2016novel}
\bibfield{author}{\bibinfo{person}{Zhiwei Jin}, \bibinfo{person}{Juan Cao},
  \bibinfo{person}{Yongdong Zhang}, \bibinfo{person}{Jianshe Zhou}, {and}
  \bibinfo{person}{Qi Tian}.} \bibinfo{year}{2016}\natexlab{b}.
\newblock \showarticletitle{Novel visual and statistical image features for
  microblogs news verification}.
\newblock \bibinfo{journal}{\emph{IEEE transactions on multimedia}}
  \bibinfo{volume}{19}, \bibinfo{number}{3} (\bibinfo{year}{2016}),
  \bibinfo{pages}{598--608}.
\newblock


\bibitem[\protect\citeauthoryear{Khattar, Goud, Gupta, and Varma}{Khattar
  et~al\mbox{.}}{2019}]%
        {khattar2019mvae}
\bibfield{author}{\bibinfo{person}{Dhruv Khattar},
  \bibinfo{person}{Jaipal~Singh Goud}, \bibinfo{person}{Manish Gupta}, {and}
  \bibinfo{person}{Vasudeva Varma}.} \bibinfo{year}{2019}\natexlab{}.
\newblock \showarticletitle{Mvae: Multimodal variational autoencoder for fake
  news detection}. In \bibinfo{booktitle}{\emph{WWW}}.
\newblock


\bibitem[\protect\citeauthoryear{Kiela, Wang, and Cho}{Kiela
  et~al\mbox{.}}{2018}]%
        {kiela2018dynamic}
\bibfield{author}{\bibinfo{person}{Douwe Kiela}, \bibinfo{person}{Changhan
  Wang}, {and} \bibinfo{person}{Kyunghyun Cho}.}
  \bibinfo{year}{2018}\natexlab{}.
\newblock \showarticletitle{Dynamic Meta-Embeddings for Improved Sentence
  Representations}. In \bibinfo{booktitle}{\emph{EMNLP}}.
  \bibinfo{pages}{1466--1477}.
\newblock


\bibitem[\protect\citeauthoryear{Kim}{Kim}{2014}]%
        {DBLP:conf/emnlp/Kim14}
\bibfield{author}{\bibinfo{person}{Yoon Kim}.} \bibinfo{year}{2014}\natexlab{}.
\newblock \showarticletitle{Convolutional Neural Networks for Sentence
  Classification}. In \bibinfo{booktitle}{\emph{EMNLP}}.
\newblock


\bibitem[\protect\citeauthoryear{Kumar and Shah}{Kumar and Shah}{2018}]%
        {kumar2018false}
\bibfield{author}{\bibinfo{person}{Srijan Kumar} {and} \bibinfo{person}{Neil
  Shah}.} \bibinfo{year}{2018}\natexlab{}.
\newblock \showarticletitle{False information on web and social media: A
  survey}.
\newblock \bibinfo{journal}{\emph{arXiv preprint arXiv:1804.08559}}
  (\bibinfo{year}{2018}).
\newblock


\bibitem[\protect\citeauthoryear{Li, Wang, and Melucci}{Li
  et~al\mbox{.}}{2019}]%
        {DBLP:conf/naacl/LiWM19}
\bibfield{author}{\bibinfo{person}{Qiuchi Li}, \bibinfo{person}{Benyou Wang},
  {and} \bibinfo{person}{Massimo Melucci}.} \bibinfo{year}{2019}\natexlab{}.
\newblock \showarticletitle{{CNM:} An Interpretable Complex-valued Network for
  Matching}. In \bibinfo{booktitle}{\emph{NAACL}}.
\newblock


\bibitem[\protect\citeauthoryear{Li, Zhao, Hu, Li, Liu, and Du}{Li
  et~al\mbox{.}}{2018}]%
        {P18-2023}
\bibfield{author}{\bibinfo{person}{Shen Li}, \bibinfo{person}{Zhe Zhao},
  \bibinfo{person}{Renfen Hu}, \bibinfo{person}{Wensi Li}, \bibinfo{person}{Tao
  Liu}, {and} \bibinfo{person}{Xiaoyong Du}.} \bibinfo{year}{2018}\natexlab{}.
\newblock \showarticletitle{Analogical Reasoning on Chinese Morphological and
  Semantic Relations}. In \bibinfo{booktitle}{\emph{ACL}}.
  \bibinfo{pages}{138--143}.
\newblock


\bibitem[\protect\citeauthoryear{Liu, Zhang, Tu, and Sun}{Liu
  et~al\mbox{.}}{2015}]%
        {liu2015statistical}
\bibfield{author}{\bibinfo{person}{Zhiyuan Liu}, \bibinfo{person}{Le Zhang},
  \bibinfo{person}{Cunchao Tu}, {and} \bibinfo{person}{Maosong Sun}.}
  \bibinfo{year}{2015}\natexlab{}.
\newblock \showarticletitle{Statistical and semantic analysis of rumors in
  chinese social media}.
\newblock \bibinfo{journal}{\emph{Scientia Sinica Informationis}}
  \bibinfo{volume}{45}, \bibinfo{number}{12} (\bibinfo{year}{2015}),
  \bibinfo{pages}{1536}.
\newblock


\bibitem[\protect\citeauthoryear{Lu, Yang, Batra, and Parikh}{Lu
  et~al\mbox{.}}{2016}]%
        {DBLP:conf/nips/LuYBP16}
\bibfield{author}{\bibinfo{person}{Jiasen Lu}, \bibinfo{person}{Jianwei Yang},
  \bibinfo{person}{Dhruv Batra}, {and} \bibinfo{person}{Devi Parikh}.}
  \bibinfo{year}{2016}\natexlab{}.
\newblock \showarticletitle{Hierarchical Question-Image Co-Attention for Visual
  Question Answering}. In \bibinfo{booktitle}{\emph{NIPS}}.
\newblock


\bibitem[\protect\citeauthoryear{Ma, Gao, Mitra, Kwon, Jansen, Wong, and
  Cha}{Ma et~al\mbox{.}}{2016}]%
        {DBLP:conf/ijcai/MaGMKJWC16}
\bibfield{author}{\bibinfo{person}{Jing Ma}, \bibinfo{person}{Wei Gao},
  \bibinfo{person}{Prasenjit Mitra}, \bibinfo{person}{Sejeong Kwon},
  \bibinfo{person}{Bernard~J. Jansen}, \bibinfo{person}{Kam{-}Fai Wong}, {and}
  \bibinfo{person}{Meeyoung Cha}.} \bibinfo{year}{2016}\natexlab{}.
\newblock \showarticletitle{Detecting Rumors from Microblogs with Recurrent
  Neural Networks}. In \bibinfo{booktitle}{\emph{IJCAI}}.
\newblock


\bibitem[\protect\citeauthoryear{Ma, Gao, and Wong}{Ma et~al\mbox{.}}{2018a}]%
        {DBLP:conf/www/MaGW18}
\bibfield{author}{\bibinfo{person}{Jing Ma}, \bibinfo{person}{Wei Gao}, {and}
  \bibinfo{person}{Kam{-}Fai Wong}.} \bibinfo{year}{2018}\natexlab{a}.
\newblock \showarticletitle{Detect Rumor and Stance Jointly by Neural
  Multi-task Learning}. In \bibinfo{booktitle}{\emph{WWW}}.
  \bibinfo{pages}{585--593}.
\newblock


\bibitem[\protect\citeauthoryear{Ma, Gao, and Wong}{Ma et~al\mbox{.}}{2018b}]%
        {DBLP:conf/acl/WongGM18}
\bibfield{author}{\bibinfo{person}{Jing Ma}, \bibinfo{person}{Wei Gao}, {and}
  \bibinfo{person}{Kam{-}Fai Wong}.} \bibinfo{year}{2018}\natexlab{b}.
\newblock \showarticletitle{Rumor Detection on Twitter with Tree-structured
  Recursive Neural Networks}. In \bibinfo{booktitle}{\emph{ACL}}.
\newblock


\bibitem[\protect\citeauthoryear{Ma, Gao, and Wong}{Ma et~al\mbox{.}}{2019}]%
        {DBLP:conf/www/Ma0W19}
\bibfield{author}{\bibinfo{person}{Jing Ma}, \bibinfo{person}{Wei Gao}, {and}
  \bibinfo{person}{Kam{-}Fai Wong}.} \bibinfo{year}{2019}\natexlab{}.
\newblock \showarticletitle{Detect Rumors on Twitter by Promoting Information
  Campaigns with Generative Adversarial Learning}. In
  \bibinfo{booktitle}{\emph{WWW}}. \bibinfo{pages}{3049--3055}.
\newblock


\bibitem[\protect\citeauthoryear{Melucci}{Melucci}{2015}]%
        {DBLP:series/irs/Melucci15}
\bibfield{author}{\bibinfo{person}{Massimo Melucci}.}
  \bibinfo{year}{2015}\natexlab{}.
\newblock \bibinfo{booktitle}{\emph{Introduction to Information Retrieval and
  Quantum Mechanics}}. \bibinfo{series}{The Information Retrieval Series},
  Vol.~\bibinfo{volume}{35}.
\newblock \bibinfo{publisher}{Springer}.
\newblock


\bibitem[\protect\citeauthoryear{Pennebaker, Boyd, Jordan, and
  Blackburn}{Pennebaker et~al\mbox{.}}{2015}]%
        {pennebaker2015development}
\bibfield{author}{\bibinfo{person}{James~W Pennebaker}, \bibinfo{person}{Ryan~L
  Boyd}, \bibinfo{person}{Kayla Jordan}, {and} \bibinfo{person}{Kate
  Blackburn}.} \bibinfo{year}{2015}\natexlab{}.
\newblock \bibinfo{booktitle}{\emph{The development and psychometric properties
  of LIWC2015}}.
\newblock \bibinfo{type}{{T}echnical {R}eport}.
\newblock


\bibitem[\protect\citeauthoryear{Qi, Cao, Yang, Guo, and Li}{Qi
  et~al\mbox{.}}{2019}]%
        {qi2019exploiting}
\bibfield{author}{\bibinfo{person}{Peng Qi}, \bibinfo{person}{Juan Cao},
  \bibinfo{person}{Tianyun Yang}, \bibinfo{person}{Junbo Guo}, {and}
  \bibinfo{person}{Jintao Li}.} \bibinfo{year}{2019}\natexlab{}.
\newblock \showarticletitle{Exploiting Multi-domain Visual Information for Fake
  News Detection}. In \bibinfo{booktitle}{\emph{ICDM}}.
\newblock


\bibitem[\protect\citeauthoryear{Qiu, Li, Li, Jiang, Hu, and Yang}{Qiu
  et~al\mbox{.}}{2018}]%
        {qiu2018revisiting}
\bibfield{author}{\bibinfo{person}{Yuanyuan Qiu}, \bibinfo{person}{Hongzheng
  Li}, \bibinfo{person}{Shen Li}, \bibinfo{person}{Yingdi Jiang},
  \bibinfo{person}{Renfen Hu}, {and} \bibinfo{person}{Lijiao Yang}.}
  \bibinfo{year}{2018}\natexlab{}.
\newblock \showarticletitle{Revisiting Correlations between Intrinsic and
  Extrinsic Evaluations of Word Embeddings}.
\newblock In \bibinfo{booktitle}{\emph{CCL}}. \bibinfo{publisher}{Springer},
  \bibinfo{pages}{209--221}.
\newblock


\bibitem[\protect\citeauthoryear{Sadrzadeh, Kartsaklis, and
  Balk{\i}r}{Sadrzadeh et~al\mbox{.}}{2018}]%
        {sadrzadeh2018sentence}
\bibfield{author}{\bibinfo{person}{Mehrnoosh Sadrzadeh},
  \bibinfo{person}{Dimitri Kartsaklis}, {and} \bibinfo{person}{Esma
  Balk{\i}r}.} \bibinfo{year}{2018}\natexlab{}.
\newblock \showarticletitle{Sentence entailment in compositional distributional
  semantics}.
\newblock \bibinfo{journal}{\emph{Annals of Mathematics and Artificial
  Intelligence}} \bibinfo{volume}{82}, \bibinfo{number}{4}
  (\bibinfo{year}{2018}), \bibinfo{pages}{189--218}.
\newblock


\bibitem[\protect\citeauthoryear{Shu, Cui, Wang, Lee, and Liu}{Shu
  et~al\mbox{.}}{2019}]%
        {DBLP:conf/kdd/ShuCW0L19}
\bibfield{author}{\bibinfo{person}{Kai Shu}, \bibinfo{person}{Limeng Cui},
  \bibinfo{person}{Suhang Wang}, \bibinfo{person}{Dongwon Lee}, {and}
  \bibinfo{person}{Huan Liu}.} \bibinfo{year}{2019}\natexlab{}.
\newblock \showarticletitle{dEFEND: Explainable Fake News Detection}. In
  \bibinfo{booktitle}{\emph{KDD}}. \bibinfo{pages}{395--405}.
\newblock


\bibitem[\protect\citeauthoryear{Shu, Mahudeswaran, Wang, Lee, and Liu}{Shu
  et~al\mbox{.}}{2018}]%
        {shu2018fakenewsnet}
\bibfield{author}{\bibinfo{person}{Kai Shu}, \bibinfo{person}{Deepak
  Mahudeswaran}, \bibinfo{person}{Suhang Wang}, \bibinfo{person}{Dongwon Lee},
  {and} \bibinfo{person}{Huan Liu}.} \bibinfo{year}{2018}\natexlab{}.
\newblock \showarticletitle{Fakenewsnet: A data repository with news content,
  social context and dynamic information for studying fake news on social
  media}.
\newblock \bibinfo{journal}{\emph{arXiv preprint arXiv:1809.01286}}
  (\bibinfo{year}{2018}).
\newblock


\bibitem[\protect\citeauthoryear{Song, Tu, Yang, Liu, and Sun}{Song
  et~al\mbox{.}}{2019}]%
        {song2018ced}
\bibfield{author}{\bibinfo{person}{Changhe Song}, \bibinfo{person}{Cunchao Tu},
  \bibinfo{person}{Cheng Yang}, \bibinfo{person}{Zhiyuan Liu}, {and}
  \bibinfo{person}{Maosong Sun}.} \bibinfo{year}{2019}\natexlab{}.
\newblock \showarticletitle{CED: Credible Early Detection of Social Media
  Rumors}.
\newblock \bibinfo{journal}{\emph{IEEE Transactions on Knowledge and Data
  Engineering}} (\bibinfo{year}{2019}).
\newblock


\bibitem[\protect\citeauthoryear{Tacchini, Ballarin, Della~Vedova, Moret, and
  de~Alfaro}{Tacchini et~al\mbox{.}}{2017}]%
        {tacchini2017some}
\bibfield{author}{\bibinfo{person}{Eugenio Tacchini}, \bibinfo{person}{Gabriele
  Ballarin}, \bibinfo{person}{Marco~L Della~Vedova}, \bibinfo{person}{Stefano
  Moret}, {and} \bibinfo{person}{Luca de Alfaro}.}
  \bibinfo{year}{2017}\natexlab{}.
\newblock \showarticletitle{Some like it hoax: Automated fake news detection in
  social networks}.
\newblock \bibinfo{journal}{\emph{arXiv preprint arXiv:1704.07506}}
  (\bibinfo{year}{2017}).
\newblock


\bibitem[\protect\citeauthoryear{Trabelsi, Bilaniuk, Zhang, Serdyuk,
  Subramanian, Santos, Mehri, Rostamzadeh, Bengio, and Pal}{Trabelsi
  et~al\mbox{.}}{2018}]%
        {DBLP:conf/iclr/TrabelsiBZSSSMR18}
\bibfield{author}{\bibinfo{person}{Chiheb Trabelsi}, \bibinfo{person}{Olexa
  Bilaniuk}, \bibinfo{person}{Ying Zhang}, \bibinfo{person}{Dmitriy Serdyuk},
  \bibinfo{person}{Sandeep Subramanian}, \bibinfo{person}{Jo{\~{a}}o~Felipe
  Santos}, \bibinfo{person}{Soroush Mehri}, \bibinfo{person}{Negar
  Rostamzadeh}, \bibinfo{person}{Yoshua Bengio}, {and}
  \bibinfo{person}{Christopher~J. Pal}.} \bibinfo{year}{2018}\natexlab{}.
\newblock \showarticletitle{Deep Complex Networks}. In
  \bibinfo{booktitle}{\emph{ICLR}}.
\newblock


\bibitem[\protect\citeauthoryear{Vaswani, Shazeer, Parmar, Uszkoreit, Jones,
  Gomez, Kaiser, and Polosukhin}{Vaswani et~al\mbox{.}}{2017}]%
        {vaswani2017attention}
\bibfield{author}{\bibinfo{person}{Ashish Vaswani}, \bibinfo{person}{Noam
  Shazeer}, \bibinfo{person}{Niki Parmar}, \bibinfo{person}{Jakob Uszkoreit},
  \bibinfo{person}{Llion Jones}, \bibinfo{person}{Aidan~N Gomez},
  \bibinfo{person}{{\L}ukasz Kaiser}, {and} \bibinfo{person}{Illia
  Polosukhin}.} \bibinfo{year}{2017}\natexlab{}.
\newblock \showarticletitle{Attention is all you need}. In
  \bibinfo{booktitle}{\emph{NIPS}}. \bibinfo{pages}{5998--6008}.
\newblock


\bibitem[\protect\citeauthoryear{Wang, Li, Melucci, and Song}{Wang
  et~al\mbox{.}}{2019}]%
        {DBLP:conf/www/WangLM019}
\bibfield{author}{\bibinfo{person}{Benyou Wang}, \bibinfo{person}{Qiuchi Li},
  \bibinfo{person}{Massimo Melucci}, {and} \bibinfo{person}{Dawei Song}.}
  \bibinfo{year}{2019}\natexlab{}.
\newblock \showarticletitle{Semantic Hilbert Space for Text Representation
  Learning}. In \bibinfo{booktitle}{\emph{WWW}}. \bibinfo{pages}{3293--3299}.
\newblock


\bibitem[\protect\citeauthoryear{Wang, Zhang, Li, Song, Hou, and Shang}{Wang
  et~al\mbox{.}}{2016}]%
        {wang2016exploration}
\bibfield{author}{\bibinfo{person}{Benyou Wang}, \bibinfo{person}{Peng Zhang},
  \bibinfo{person}{Jingfei Li}, \bibinfo{person}{Dawei Song},
  \bibinfo{person}{Yuexian Hou}, {and} \bibinfo{person}{Zhenguo Shang}.}
  \bibinfo{year}{2016}\natexlab{}.
\newblock \showarticletitle{Exploration of quantum interference in document
  relevance judgement discrepancy}.
\newblock \bibinfo{journal}{\emph{Entropy}} \bibinfo{volume}{18},
  \bibinfo{number}{4} (\bibinfo{year}{2016}), \bibinfo{pages}{144}.
\newblock


\bibitem[\protect\citeauthoryear{Wang, Ma, Jin, Yuan, Xun, Jha, Su, and
  Gao}{Wang et~al\mbox{.}}{2018}]%
        {DBLP:conf/kdd/WangMJYXJSG18}
\bibfield{author}{\bibinfo{person}{Yaqing Wang}, \bibinfo{person}{Fenglong Ma},
  \bibinfo{person}{Zhiwei Jin}, \bibinfo{person}{Ye Yuan},
  \bibinfo{person}{Guangxu Xun}, \bibinfo{person}{Kishlay Jha},
  \bibinfo{person}{Lu Su}, {and} \bibinfo{person}{Jing Gao}.}
  \bibinfo{year}{2018}\natexlab{}.
\newblock \showarticletitle{{EANN:} Event Adversarial Neural Networks for
  Multi-Modal Fake News Detection}. In \bibinfo{booktitle}{\emph{KDD}}.
  \bibinfo{pages}{849--857}.
\newblock


\bibitem[\protect\citeauthoryear{Wenzel}{Wenzel}{2019}]%
        {wenzel2019verify}
\bibfield{author}{\bibinfo{person}{Andrea Wenzel}.}
  \bibinfo{year}{2019}\natexlab{}.
\newblock \showarticletitle{To verify or to disengage: Coping with “fake
  news” and ambiguity}.
\newblock \bibinfo{journal}{\emph{International Journal of Communication}}
  \bibinfo{volume}{13} (\bibinfo{year}{2019}), \bibinfo{pages}{19}.
\newblock


\bibitem[\protect\citeauthoryear{Wu, Li, Hu, and Liu}{Wu et~al\mbox{.}}{2017}]%
        {DBLP:conf/sdm/WuLHL17}
\bibfield{author}{\bibinfo{person}{Liang Wu}, \bibinfo{person}{Jundong Li},
  \bibinfo{person}{Xia Hu}, {and} \bibinfo{person}{Huan Liu}.}
  \bibinfo{year}{2017}\natexlab{}.
\newblock \showarticletitle{Gleaning Wisdom from the Past: Early Detection of
  Emerging Rumors in Social Media}. In \bibinfo{booktitle}{\emph{SDM}}.
  \bibinfo{pages}{99--107}.
\newblock


\bibitem[\protect\citeauthoryear{Yang, Ma, Li, Tsai, and Salakhutdinov}{Yang
  et~al\mbox{.}}{2020b}]%
        {yang2019complex}
\bibfield{author}{\bibinfo{person}{Muqiao Yang}, \bibinfo{person}{Martin~Q Ma},
  \bibinfo{person}{Dongyu Li}, \bibinfo{person}{Yao-Hung~Hubert Tsai}, {and}
  \bibinfo{person}{Ruslan Salakhutdinov}.} \bibinfo{year}{2020}\natexlab{b}.
\newblock \showarticletitle{Complex Transformer: A Framework for Modeling
  Complex-Valued Sequence}. In \bibinfo{booktitle}{\emph{ICASSP}}.
  \bibinfo{pages}{4232--4236}.
\newblock


\bibitem[\protect\citeauthoryear{Yang, Lyu, Tian, Liu, Liu, and Zhang}{Yang
  et~al\mbox{.}}{2020a}]%
        {DBLP:conf/ijcai/YangLTLLZ20}
\bibfield{author}{\bibinfo{person}{Xiaoyu Yang}, \bibinfo{person}{Yuefei Lyu},
  \bibinfo{person}{Tian Tian}, \bibinfo{person}{Yifei Liu},
  \bibinfo{person}{Yudong Liu}, {and} \bibinfo{person}{Xi Zhang}.}
  \bibinfo{year}{2020}\natexlab{a}.
\newblock \showarticletitle{Rumor Detection on Social Media with Graph
  Structured Adversarial Learning}. In \bibinfo{booktitle}{\emph{IJCAI}}.
  \bibinfo{pages}{1417--1423}.
\newblock


\bibitem[\protect\citeauthoryear{Yang, Yang, Dyer, He, Smola, and Hovy}{Yang
  et~al\mbox{.}}{2016}]%
        {DBLP:conf/naacl/YangYDHSH16}
\bibfield{author}{\bibinfo{person}{Zichao Yang}, \bibinfo{person}{Diyi Yang},
  \bibinfo{person}{Chris Dyer}, \bibinfo{person}{Xiaodong He},
  \bibinfo{person}{Alexander~J. Smola}, {and} \bibinfo{person}{Eduard~H.
  Hovy}.} \bibinfo{year}{2016}\natexlab{}.
\newblock \showarticletitle{Hierarchical Attention Networks for Document
  Classification}. In \bibinfo{booktitle}{\emph{NAACL-HLT}}.
  \bibinfo{pages}{1480--1489}.
\newblock


\bibitem[\protect\citeauthoryear{Ying, Zhuang, Zhang, Liu, Xu, Xie, Xiong, and
  Wu}{Ying et~al\mbox{.}}{2018}]%
        {ying2018sequential}
\bibfield{author}{\bibinfo{person}{Haochao Ying}, \bibinfo{person}{Fuzhen
  Zhuang}, \bibinfo{person}{Fuzheng Zhang}, \bibinfo{person}{Yanchi Liu},
  \bibinfo{person}{Guandong Xu}, \bibinfo{person}{Xing Xie},
  \bibinfo{person}{Hui Xiong}, {and} \bibinfo{person}{Jian Wu}.}
  \bibinfo{year}{2018}\natexlab{}.
\newblock \showarticletitle{Sequential recommender system based on hierarchical
  attention network}. In \bibinfo{booktitle}{\emph{IJCAI}}.
\newblock


\bibitem[\protect\citeauthoryear{Yu, Liu, Wu, Wang, and Tan}{Yu
  et~al\mbox{.}}{2017}]%
        {ijcai2017-545}
\bibfield{author}{\bibinfo{person}{Feng Yu}, \bibinfo{person}{Qiang Liu},
  \bibinfo{person}{Shu Wu}, \bibinfo{person}{Liang Wang}, {and}
  \bibinfo{person}{Tieniu Tan}.} \bibinfo{year}{2017}\natexlab{}.
\newblock \showarticletitle{A Convolutional Approach for Misinformation
  Identification}. In \bibinfo{booktitle}{\emph{IJCAI}}.
\newblock


\end{thebibliography}

\appendix

\end{document}